\documentclass[%
 reprint,
 superscriptaddress,
amsmath,amssymb,
aps,
pra,
floatfix,
]{revtex4-2}

\usepackage{printlen}
\uselengthunit{in}  % Set unit to inches
\usepackage{graphicx}% Include figure files
\usepackage{dcolumn}% Align table columns on decimal point
\usepackage{xcolor}
\usepackage{bm}% bold math
\usepackage{hyperref}% add hypertext capabilities
\usepackage{braket}
\begin{document}

\title{Temporal Pulse Origins in Atom Interferometric Quantum Sensors}

\author{Jack Saywell}
    \email{j.c.saywell@soton.ac.uk}
\author{Nikolaos Dedes}
\thanks{Now at Thales UK}
\author{Max Carey}
    \affiliation{\mbox{School of Physics and Astronomy, University of Southampton, Highfield, Southampton SO17 1BJ, United Kingdom}}
\author{Brynle Barrett}
    \affiliation{\mbox{Department of Physics, University of New Brunswick, 8 Bailey Dr., Fredericton, New Brunswick E3B 5A3, Canada}}
\author{Tim Freegarde}
    \affiliation{\mbox{School of Physics and Astronomy, University of Southampton, Highfield, Southampton SO17 1BJ, United Kingdom}}

\date{\today}

\begin{abstract}
Quantum sensors based upon atom interferometry typically rely on radio-frequency or optical pulses to coherently manipulate atomic states and make precise measurements of inertial and gravitational effects. An advantage of these sensors over their classical counterparts is often said to be that their measurement scale factor is precisely known and highly stable. However, in practice the finite pulse duration makes the sensor scale factor dependent upon the pulse shape and sensitive to variations in control field intensity, frequency, and atomic velocity. Here, we explore the concept of a temporal pulse origin in atom interferometry, where the inertial phase response of any pulse can be parameterized using a single point in time. We show that the temporal origin permits a simple determination of the measurement scale factor and its stability against environmental perturbations. Moreover, the temporal origin can be treated as a tunable parameter in the design of tailored sequences of shaped pulses to enhance scale factor stability and minimize systematic errors. We demonstrate through simulations that this approach to pulse design can reduce overall sequence durations while increasing robustness to realistic fluctuations in control field amplitude. Our results show that the temporal pulse origin explains a broad class of systematic errors in existing devices and enables the design of short, robust pulses which we expect will improve the performance of current and next-generation interferometric quantum sensors.
\end{abstract}

\maketitle

\section{\label{sec:1}Introduction}

Quantum technologies such as atom-interferometric sensors~\cite{kasevich1991atomic}, solid-state quantum magnetometers~\cite{balasubramanian2008nanoscale}, and quantum information processors~\cite{divincenzo2000physical} all require high-fidelity coherent control of quantum states to reach their potential performance beyond proof-of-principle demonstrations. However, these operations are inherently sensitive to inhomogeneities in the control field and environment. Consequently, much effort has been spent both attempting to eliminate these inhomogeneities entirely - typically at the expense of experimental simplicity - and mitigate their impact on performance through techniques such as quantum error correction~\cite{steane1996error} and robust quantum control~\cite{glaser2015training}.

In light-pulse atom interferometers - which may be configured as highly-precise inertial~\cite{barrett2014sagnac, canuel2006six, d2024atom, templier2022tracking, geiger2011detecting} and gravitational sensors~\cite{kasevich1991atomic, mcguirk2002sensitive, bidel2018absolute, stray2022quantum} with applications ranging from navigation~\cite{narducci2022advances, phillips2022position} to fundamental physics~\cite{badurina2020aion, canuel2018exploring, hamilton2015atom} - sequences of optical or radio-frequency (RF) $\pi/2$ ``beamsplitter'' and $\pi$ ``mirror'' pulses are used to coherently manipulate atomic matter-waves. A key advantage of these sensors over classical devices is that in principle their measurement scale factor - defined as the output signal per unit quantity of interest - is both well known and highly stable. In practice, however, the finite duration and shape of the optical pulses modify the scale factor~\cite{fang2018improving, templier2022tracking, bonnin2015characterization} and can introduce an unwanted sensitivity to atomic velocity~\cite{gillot2016limits}, optical phase noise~\cite{fang2018improving}, platform motion~\cite{barrett2016dual, d2024atom}, and laser intensity fluctuations~\cite{dedes2023optimizing}. This reduces the fidelity of conventionally-employed simple ``rectangular'' or Gaussian-shaped interferometer pulses, lowering the sensor's signal-to-noise ratio (SNR)~\cite{szigeti2012momentum}. 

One solution is to replace conventional pulses and sequences with alternatives designed using quantum optimal control to be robust to inhomogeneities such as laser intensity noise and/or increase SNR by accepting atomic sources with broader velocity distributions~\cite{saywell2018optimal, saywell2020optimal, dedes2023optimizing, louie2023robust, chen2023enhancing, saywell2020biselective, lahuerta2025diffraction}. This approach was recently successfully employed experimentally to suppress laser intensity noise in a Bragg pulse atom interferometer~\cite{saywell2023enhancing} and enhance the fidelity of momentum transfer in large-area interferometry~\cite{rodzinka2024optimal}. Little attention has been paid, however, to the effect of optimized pulses or sequences on the interferometer's scale factor. Additionally, the duration of individually-optimized pulses is typically many times longer than conventional pulses, inevitably causing signal loss from spontaneous emission in extended sequences~\cite{kotru2015large} and thereby outweighing any potential gain in fidelity.
    
In this paper, we address these problems by exploring the concept of a temporal pulse origin in atom-interferometric quantum sensors. We show that the inertial phase response of any finite-duration pulse may be parameterized by a single point in time: its temporal origin. The temporal origins of all pulses in a sequence determine the measurement scale factor, thereby providing a framework to assess scale factor stability: stable pulse origins imply a stable scale factor and vice versa. We also show how the pulse origin can explain systematic effects caused by imperfect compensation of the acceleration-induced Doppler shift through adjustments to laser frequency and phase~\cite{templier2022tracking}.

Furthermore, the pulse origin itself can be included as a parameter in the optimization of individual pulses and entire sequences. For beamsplitter pulses, we show that this approach can significantly reduce pulse durations without appreciable loss of robustness or fidelity. Finally, we demonstrate through numerical simulations that beamsplitter pulses designed with stable pulse origins enhance scale factor stability and maintain interferometer symmetry in the presence of realistic fluctuations in laser intensity. We show that this approach can reduce the maximum scale factor error in a $T=5\ \mathrm{ms}$ interferometer by $21\times$ in the presence of $10\%$ intensity fluctuations at the expense of only $\pi$ radians of additional pulse area (i.e. a half Rabi cycle) per pulse, over that of simple rectangular pulses. 

\begin{figure}[!tb]
\includegraphics[width=\columnwidth]{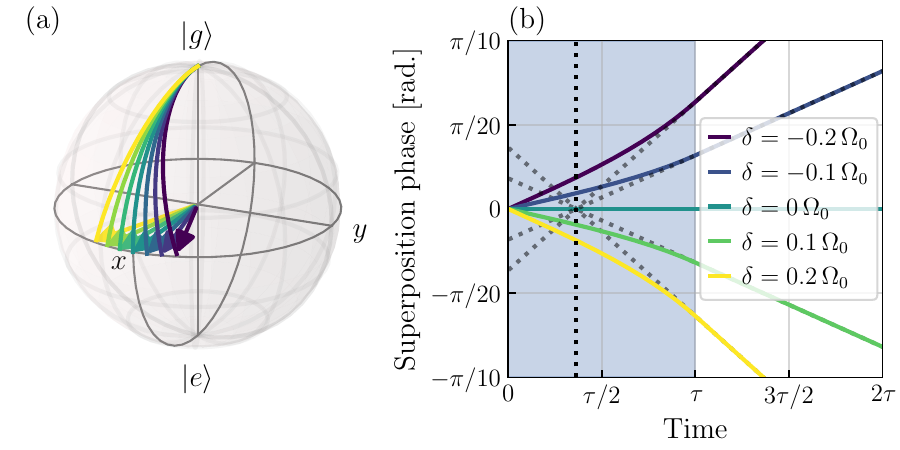}
\caption{\label{fig:origin_pi2} (a) Bloch sphere trajectories for a range of detunings during a rectangular $\pi/2$ pulse. (b) The superposition phase during (blue shaded region) and immediately after a rectangular $\pi/2$ pulse is depicted for a range of detunings. The non-zero phase dispersion causes differently detuned atoms to accumulate different phases during the pulse. Since these phases depend linearly on detuning, they can all be traced back to a common origin in time.}
\end{figure}

\section{Temporal pulse origin}

\begin{figure*}[thb]
  \centering
  \begin{minipage}[b]{0.48\textwidth}
    \includegraphics[width=\linewidth]{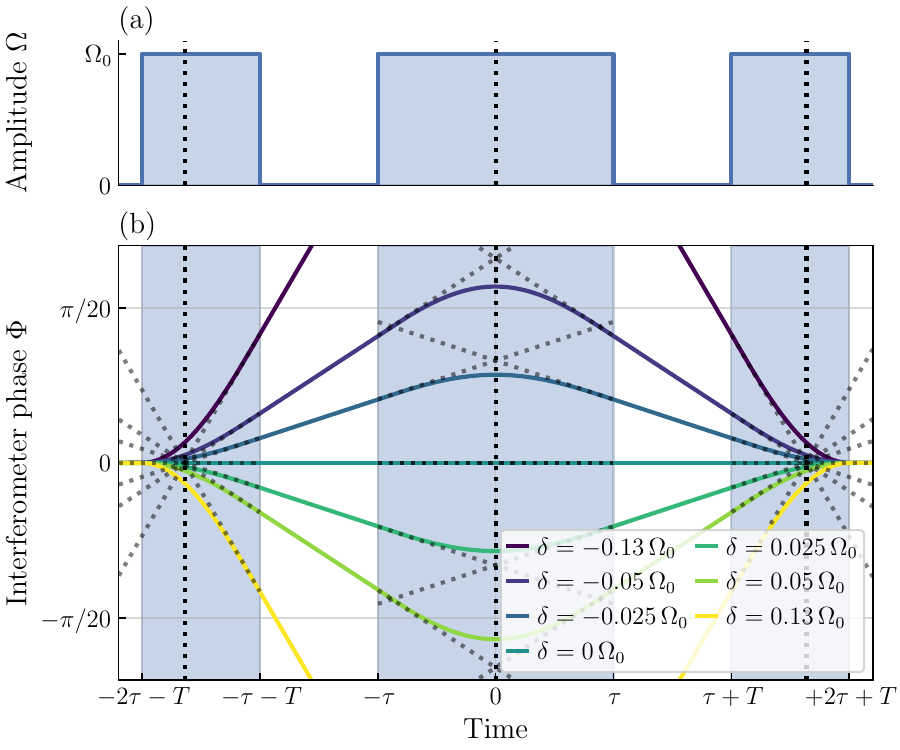}
  \end{minipage}%
    \hspace{0.02\linewidth}  % Small horizontal space
  \begin{minipage}[b]{0.48\textwidth}
    \includegraphics[width=\linewidth]{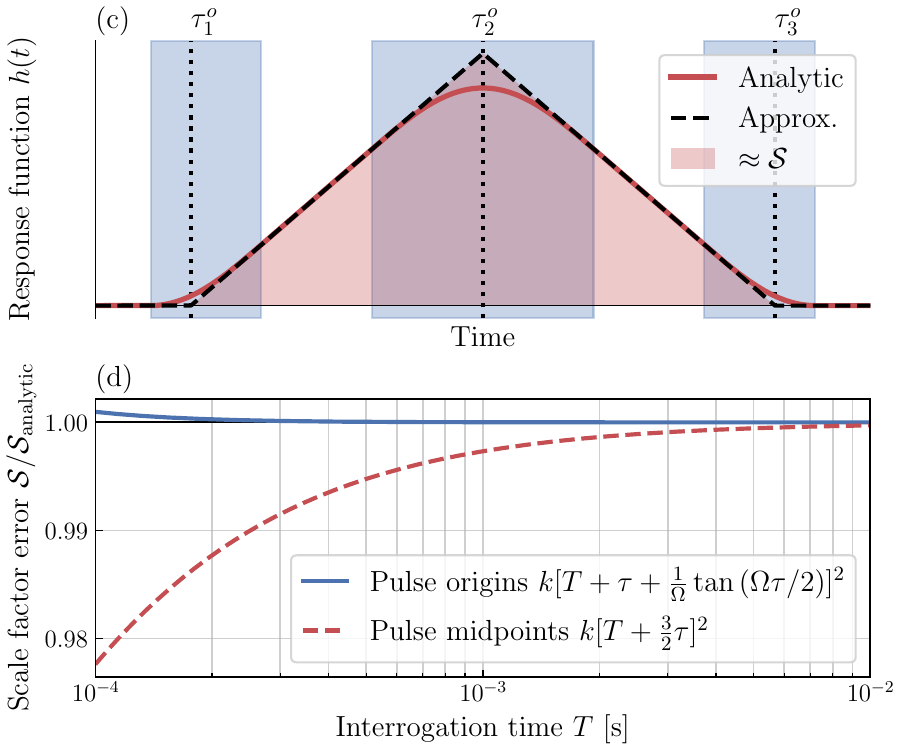}
  \end{minipage}
  \caption{\label{fig:origin_diagram} 
  The amplitude profile (a) and interferometer phase (b) as functions of time during a three-pulse Mach-Zehnder sequence comprised of rectangular $\pi/2$ (beamsplitter) and $\pi$ (mirror) pulses. The interferometer phase is shown for a range of detunings near resonance. The diagonal dotted lines show how the change in phase due to each pulse can be traced back to a common point in time - the pulse origin. (c) shows the response function for a Mach-Zehnder sequence composed of rectangular pulses. The scale factor is given by the area under the response function, which may be approximated as triangular in the limit where $T\gg\tau$. (d) shows the fractional scale factor error $\mathcal{S}/\mathcal{S}_{\mathrm{analytic}}$ for the same sequence as a function of the interrogation time $T$. $\mathcal{S}_{\mathrm{analytic}}=k\int_{-\infty}^{+\infty} h(t)\, \mathrm{d}t$ is computed using the analytic expression for $h(t)$ from~\cite{bonnin2015characterization}. The dashed red and solid blue lines correspond to the scale factor obtained by approximating the response function as a triangle whose vertices occur at the temporal centers and temporal origins of each pulse, respectively. Using the pulse origins results in a more accurate value for the scale factor. The beamsplitter pulse duration $\tau$ was 10~$\mu$s. The Rabi frequency was held constant for all pulses.
  }
  \label{fig:both}
\end{figure*}

Without loss of generality, we introduce the temporal pulse origin in the context of a three-pulse Mach-Zehnder sequence consisting of an initial $\pi/2$ “beamsplitter” pulse, a central $\pi$ “mirror” pulse, and a final $\pi/2$ beamsplitter (or ``recombiner'') pulse each separated by periods of free evolution. We label these three pulses $1$, $2$, and $3$, respectively. For velocity-selective transitions, the interferometer phase $\Phi$ following this sequence depends on the atomic acceleration relative to the laser wave-fronts~\cite{bongs2006high}, making it suitable for inertial and gravitational sensing.

In the Bloch sphere picture~\cite{feynman1957geometrical} (described further in Appendix~\ref{appendix:atom-light}), the action of each pulse corresponds to a rotation of the atomic state vector. The poles of the Bloch sphere represent the basis states ($\ket{g,e}$) of the two primary interferometric ``arms" and all other surface points are superposition states with variable relative phase (differing in longitude) and amplitude (differing in latitude). Resonant pulses with constant phase and frequency cause rotations of the state vector about axes in the equatorial plane at a rate given by the Rabi frequency $\Omega$, which we refer to as the pulse amplitude. On resonance, beamsplitter pulses correspond to rotations by $90^{\circ}$, and mirror pulses by $180^{\circ}$. As the atomic detuning $\delta$ increases from resonance, the rotation axis tilts out of the equatorial plane, deflecting trajectories from their ideal path and reducing pulse fidelity.

The sensitivity of an atom interferometer’s phase to external influences evolves during each $\pi/2$ or $\pi$ pulse. Figure~\ref{fig:origin_pi2} shows the example of the detuning dependence of a rectangular $\pi/2$ pulse of duration $\tau=\pi/(2\Omega_0)$, where $\Omega_0$ is the nominal Rabi frequency: all states begin at the pole, but progressively fan out during their passage towards the equator. The Bloch sphere trajectories for different detunings are shown in Figure~\ref{fig:origin_pi2}~(a), and the atomic superposition phases are shown in the shaded region of Figure~\ref{fig:origin_pi2}~(b). Close to resonance (where $\delta$ is small in comparison with the Rabi frequency $\Omega$), the superposition phases have a common temporal dependence, scaled linearly with $\delta$. At the end of the pulse, the superposition phase is therefore a linear function of detuning, and the freely-evolving atomic states subsequently rotate about the Bloch sphere $z$-axis with an angular velocity $\delta$. The interferometer phase trajectories in this period are therefore the same as those for an infinitely short $\pi/2$ pulse occurring at time $\tau^o=\tau + \partial \Phi/\partial \delta$, where $\Phi = f(\delta)$ is the phase dispersion at the end of the pulse. We shall refer to $\tau^o$ as the temporal origin of the pulse. For a rectangular beamsplitter pulse with Rabi frequency $\Omega$, we find that
\begin{equation}\label{eq:rectangular-origin}
    \tau_1^o=\tau-\frac{1}{\Omega}\tan(\Omega\tau/2).
\end{equation}
Importantly, the pulse origin for the rectangular $\pi/2$ pulse depends on the Rabi frequency and hence also on the laser intensity. It is this apparent origin, which depends upon the pulse shape in amplitude and phase, that determines the scale factor for an interferometric measurement.

\subsection{Rectangular pulse sequences}

We can apply the pulse origin concept to a full sequence of three rectangular pulses by considering how the interferometer phase varies throughout the sequence. This can be done using the sensitivity function formalism~\cite{dick1989local, cheinet2008measurement,fang2018improving}. The phase sensitivity function $g(t)$ describes the response of the sensor to a step-like perturbation in the phase $\delta\phi(t)$ of the laser driving the transition:
\begin{equation}
    g(t)=\lim\limits_{\delta\phi \rightarrow0}\frac{\delta\Phi(\delta\phi, t)}{\delta\phi(t)}.
\end{equation}
Using $g(t)$, we find that the interferometer phase is
\begin{equation}
    \Phi(\delta,t)=\int_{-\infty}^{t} g(t')\frac{\mathrm{d}\phi(t')}{\mathrm{d}t'}\, \mathrm{d}t'=\int_{-\infty}^{t} g(t')\, \delta\, \mathrm{d}t'.
\end{equation}
Figure~\ref{fig:origin_diagram} (b) depicts the phase throughout an entire Mach-Zehnder sequence of rectangular pulses for differently detuned atoms in the absence of acceleration. In this case, since the Rabi frequency of each pulse is the same, the detuning-dependent phases exactly cancel at the end of the sequence, forming a perfect spin echo. If the phases cancel, the interferometer is said to be ``closed" meaning there is zero sensitivity to initial atomic velocity along the measurement axis, which we take to be along $z$.

We can also see from Figure~\ref{fig:origin_diagram} that the temporal origin of the rectangular mirror pulse, defined by the crossing point of the free-evolution phases extrapolated during the pulse, coincides with its temporal mid-point. This means that a detuned rectangular $\pi$ pulse contributes no phase error to the interferometer. Similarly, we see from symmetry that the origin of the final beamsplitter pulse is given by $({1}/{\Omega})\tan(\Omega\tau/2)$ relative to the pulse start time. If the Rabi frequencies of the first and final pulses differ, this symmetry is broken; the temporal origins shift in time and the interferometer becomes sensitive to atomic velocity along $z$~\cite{gillot2016limits}.

The pulse origins enable one to obtain the measurement scale factor of the interferometer, which is given by $\mathcal{S}=\partial\Phi/\partial a$, where $a$ is the measured property - in this case the atomic acceleration (assumed constant) along the optical wave-vector direction $\mathbf{k}$. This is given by the area under the interferometer's acceleration response function~\cite{bonnin2015characterization} $h(t)$, which is depicted in Figure~\ref{fig:origin_diagram} (c). $h(t)$ is defined as follows
\begin{equation}\label{eq:response}
   h(t)=\int_t^{+\infty}g(t')\,\mathrm{d}t'.
\end{equation}

In the case where $T\gg\tau$, the response function is approximately triangular with vertices located at the three temporal origins rather than at the centers of each pulse as one might expect. In the case of a closed interferometer where all pulses operate with the same Rabi frequency, the area under this triangle therefore yields an approximate value for the interferometer scale factor:
\begin{equation}
    \mathcal{S}\approx k\bigg[T+\tau+\frac{1}{\Omega}\tan(\Omega \tau/2)\bigg]^2.
\end{equation}
This reproduces the well-known result from~\cite{templier2022tracking, templier2021three, bonnin2015characterization} when comparing only those terms that depend on pulse durations to first-order. We note that this improved triangular approximation was also made in Ref.~\cite{templier2022tracking} to enhance the correlations within a hybrid accelerometer using rectangular pulses however the vertices were not identified with the temporal origins as described here. We also note that, in the rectangular case, the central $\pi$ pulse can also be considered as two back-to-back $\pi/2$ pulses with individual origins. This leads to an improved approximation of the response function, which we discuss further in Appendix~\ref{appendix:mirror-dead-time}.

Figure~\ref{fig:origin_diagram} (d) shows a comparison between two different methods of calculating the scale factor for a rectangular pulse interferometer: integrating the triangular approximation of $h(t)$ using the pulse midpoints as vertices and the triangular approximation using the pulse origins. The beamsplitter duration was $\tau=10\ \mu$s. Using the triangular approximation with the pulse origins rather than the pulse mid-points provides a more accurate value for the scale factor as the interrogation time is reduced. In the next section, we extend this result to the case of arbitrary shaped pulses.

\subsection{Shaped pulse sequences}\label{sec:shaped-pulses}

\begin{figure}[tb]
\includegraphics[width=\columnwidth]{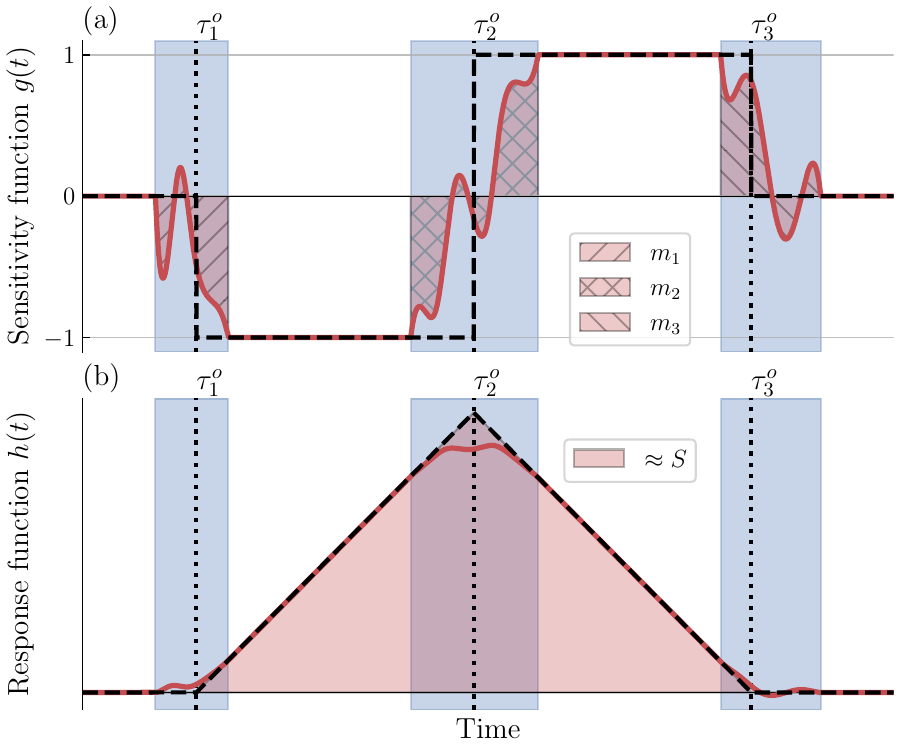}
\caption{\label{fig:origin_diagram_shaped} 
 (a) and (b) show the phase sensitivity and inertial response functions for a three-pulse Mach-Zehnder sequence composed of arbitrary shaped pulses. For small detunings, the area under the sensitivity function during each pulse is given by the gradient $m_1$, $m_2$ or $m_3$ of the phase dispersion for that pulse near resonance. The interferometer scale factor $\mathcal{S}$ - given by the area under the response function - can be approximated by considering all three pulses to occur instantaneously at their respective temporal origins ($\tau_1^o, \tau_2^o, \tau_3^o$) - shown by the solid dashed curves in (a) and (b). The precise shape of the sensitivity function shown within each pulse is arbitrary.
}
\end{figure}

We now generalize to Mach-Zehnder sequences composed of three arbitrarily shaped pulses $1,2,3$ with durations $\tau_j,\ j=1,2,3$. For a shaped pulse, parameters such as the Rabi frequency (amplitude) and laser control phase can be time-dependent. We are free to split the total interferometer phase into contributions from each section of the interferometer as follows:
\begin{equation}
    \Phi=\Phi_1+\Phi_{T1}+\Phi_2+\Phi_{T2}+\Phi_3.
\end{equation}
$\Phi_j$ is the contribution from pulse $j$ (evaluated at the final time of the pulse). $\Phi_{T1,T2}$ are the free evolution phases. Using the sensitivity function, we find:
\begin{subequations}
\begin{align}
\Phi_1=&\int_{t_i}^{t_i+\tau_1}g(t)\frac{\mathrm{d}\phi(t)}{\mathrm{d}t}\,\mathrm{d}t, \\
\Phi_2=&\int_{t_i+\tau_1+T}^{t_i+\tau_1+\tau_2+T}g(t)\frac{\mathrm{d}\phi(t)}{\mathrm{d}t}\,\mathrm{d}t, \\
\Phi_3=&\int_{t_i+\tau_1+\tau_2+2T}^{t_i+\tau_1+\tau_2+\tau_3+2T}g(t)\frac{\mathrm{d}\phi(t)}{\mathrm{d}t}\,\mathrm{d}t.
\end{align}
\end{subequations}
$t_i$ is the starting time of the pulse sequence.

If we assume that the change in velocity-sensitive detuning $\delta=\mathrm{d}\phi/\mathrm{d}t$ during each pulse is much smaller than its average value, then we can rewrite the above in terms of the gradient of the phase dispersion of each pulse $m_j
\equiv\partial\Phi_j/\partial\delta$ as
\begin{subequations}\label{eq:misens}
\begin{align}\label{eq:m1sens}
m_1\approx&\int_{t_i}^{t_i+\tau_1}g(t)\,\mathrm{d}t, \\ \label{eq:m2sens}
m_2\approx&\int_{t_i+\tau_1+T}^{t_i+\tau_1+\tau_2+T}g(t)\,\mathrm{d}t, \\ \label{eq:m3sens}
m_3\approx&\int_{t_i+\tau_1+\tau_2+2T}^{t_i+\tau_1+\tau_2+\tau_3+2T}g(t)\,\mathrm{d}t.
\end{align}
\end{subequations}
This means that the gradient of the phase dispersion for each pulse is given by the area under the sensitivity function during that pulse, as depicted in Figure~\ref{fig:origin_diagram_shaped} (a). We also note that we have adopted a sign convention where $m_1$ is negative. Equations~\eqref{eq:misens} are valid close to resonance. Each $m_i$ has units of time.

For the interferometer to remain insensitive to the initial atomic velocity, the total area under the sensitivity function must be zero. In the case of equal free-evolution periods we see that this requirement means that $m_1+m_2+m_3=0$. In other words, it is possible to engineer an interferometric accelerometer composed of pulses with variable $m_j$ with the requirement that they sum to zero over the entire sequence. 

Using Equation~\eqref{eq:response}, and recalling that the response function must be zero at the end of the interferometer, we can rewrite Equations~\eqref{eq:misens} in terms of the response function $h(t)$:
\begin{subequations}
\begin{align}
    m_1 &\approx h(t_i)-h(t_i+\tau_1), \\
    m_2 &\approx h(t_i+\tau_1+T)-h(t_i+\tau_1+\tau_2+T), \\
    m_3 &\approx h(t_i+\tau_1+\tau_2+2T).
\end{align}
\end{subequations}
The gradient of the phase dispersion is therefore approximately given by the change in the response function during the pulse. Extrapolating the linear response function during free evolution into each finite pulse and assuming the interferometer remains closed ($h(t_i)=0$), we obtain a triangle, as shown in Figure~\ref{fig:origin_diagram_shaped} (b). The temporal pulse origins are given by the vertices of this triangle.

In general, we can relate the gradients $m_j$ to the temporal origins of each pulse as follows:
\begin{subequations}
\begin{align}\label{eq:bs-origin}
\tau^o_1&=\tau_1+m_1, \\
\tau^o_2&=(\tau_2-m_2)/2, \\
\tau^o_3&=m_3.
\end{align}
\end{subequations}
Each origin is measured from the start of the respective pulse. A shift in the origin of any one pulse will ``open'' the interferometer unless balanced by a corresponding shift in one or more of the remaining pulses. Put differently, the interferometer will remain closed and insensitive to velocity if the pulse origins are equally spaced. 

We can also draw a link between the shape of the pulse (which determines its sensitivity function) and its temporal origin. For example, in the case of a mirror pulse with symmetric amplitude profile $\Omega(t)$, and anti-symmetric phase profile $\phi(t)$, the sensitivity function must integrate to zero during the pulse~\cite{saywell2020optimal}. This means its origin will lie at the center and, importantly, remain there unless the pulse symmetry is broken e.g. due to distortions from imperfect experimental realization of the waveform.

\begin{figure*}[t]
\includegraphics[width=\linewidth]{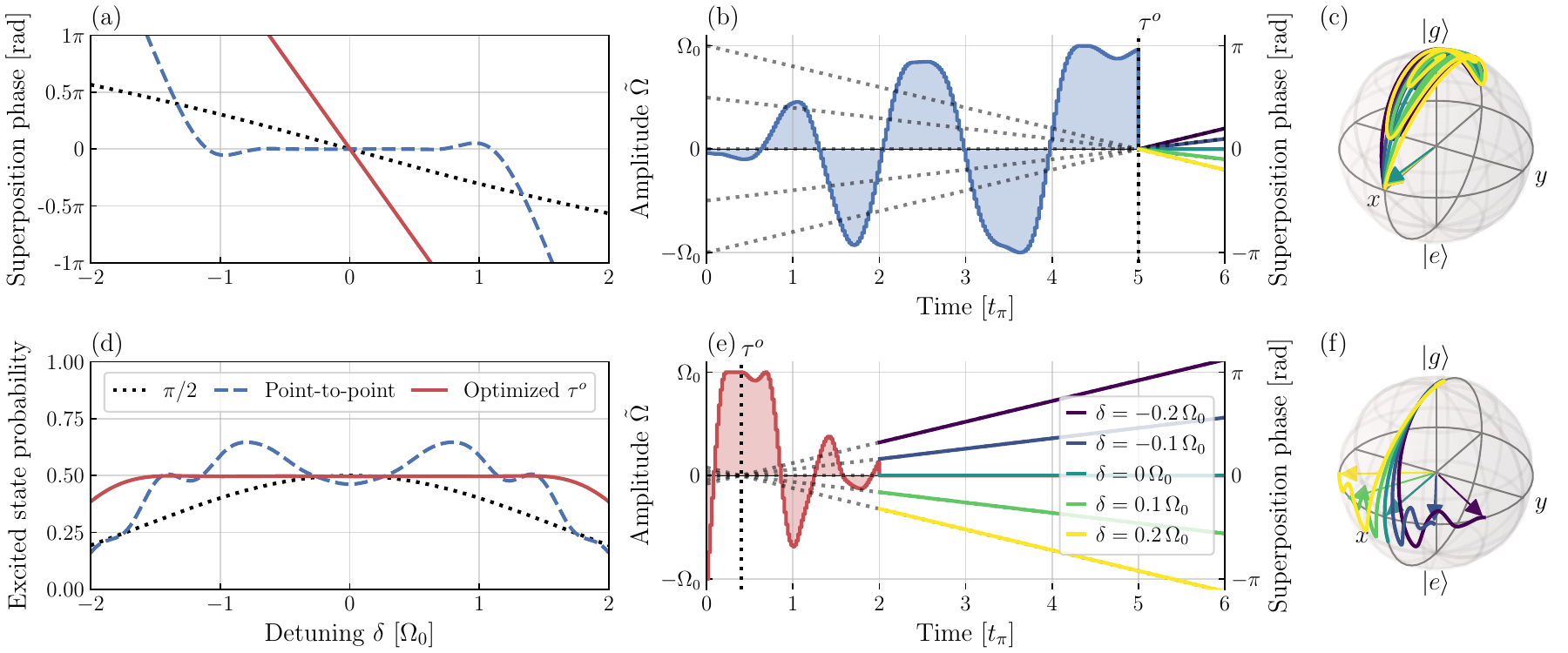}
\caption{\label{fig:optimised_pulse_comparison} The superposition phase (a) and excited state probability (d) as functions of detuning after three different beamsplitter pulses: rectangular $\pi/2$, point-to-point ($\tau^o=\tau$) and optimized origin ($\tau^o<\tau$). (b) and (e) show the amplitude profiles for each optimized pulse overlayed with the superposition phases computed as a function of time immediately after each pulse for a range of detunings (in units of $\Omega_0$). The pulse origins are found for each pulse by tracing back the superposition phases accumulated during each pulse (grey dotted lines). (c) and (f) show Bloch sphere trajectories during each optimized pulse for the same range of detuning. The optimized pulse in (b) refocuses all atomic spins to a single point on the equator of the Bloch sphere, while the shorter pulse (e) places them on the equator with a linear spread in phase.}
\end{figure*}

For the purposes of calculating the sensor's scale factor we can approximate both $g(t)$ and $h(t)$ as if the pulses are instantaneous occurring at their respective origins. The scale factor is therefore approximately given by
\begin{equation}\label{eq:approx-scale-factor}
    \mathcal{S} \approx\frac{k}{2}\big[[T+(\tau_1-\tau^o_1)+\tau^o_2]^2 + [T+(\tau_2-\tau^o_2)+\tau^o_3]^2\big].
\end{equation}
Expanding and keeping only those terms that depend upon $T$ (including terms that are first-order in pulse durations only), this simplifies further to
\begin{equation}
    \mathcal{S} \approx k(T^2+T\tau_2-m_1T+m_3T).
\end{equation}
In Appendix~\ref{appendix:open-scale-factor} we show that Equation~\eqref{eq:approx-scale-factor} also holds if the interferometer does not close and discuss the importance of the definition of initial velocity when calculating the scale factor. 

These results show that the temporal pulse origin is a simple alternative approach to calculate the interferometer scale factor in the case of arbitrary finite-duration pulses. Moreover, it explains how these corrections can be viewed as arising from the acceleration-induced change in resonance between pulses in the interferometer. We also see that if the pulse origins vary, e.g. with laser intensity, then so too may the interferometer's scale factor.

\section{Pulse origin as a design parameter}

Viewed in terms of pulse origins, most previous approaches to shaped pulse optimization in interferometry have designed beamsplitters with the origin placed at the end of the pulse~\cite{dedes2023optimizing, saywell2021can}; and mirrors with the origin coincident with the temporal midpoint~\cite{saywell2023enhancing, saywell2020optimal, saywell2022efficient, louie2023robust}. Although fixing the temporal origins at these points in time can suppress the effect of light shifts~\cite{dedes2023optimizing}, these systematics can also be mitigated through calibration and/or wave-vector reversal, making it interesting to consider the case where the origins are placed at different locations within each pulse. 

We adapt Gradient Ascent Pulse Engineering (GRAPE)~\cite{khaneja2005optimal} to design shaped beamsplitter pulses for atom interferometry. An overview of our approach and an explanation of how the pulse origin may be included as a design parameter are provided below; a detailed description of our optimization algorithm can be found in Ref.~\cite{saywell2022efficient}.

For simplicity, in this work we design piecewise-constant shaped pulses parameterized by a single experimental control vector $\{\tilde{\Omega}^n\}=\{\Omega^n\cos(\phi^n)\}$. The laser phase is restricted to integer multiples of $\pi$ and the index $n=\{1,\ \ldots,\ M\}$ runs over the $M$ time-slices that we take to have equal duration $\Delta t=400$~ns. Starting from a random initial guess for the waveform we evaluate a figure of merit $\mathcal{F}$ for the pulse. By calculating the gradient of $\mathcal{F}$ with respect to the vector $\tilde{\Omega}^n$, and approximating the Hessian, we apply the quasi-Newton limited-memory Broyden–Fletcher–Goldfarb–Shanno (L-BFGS) method~\cite{de2011second} to optimize the waveform and maximize $\mathcal{F}$. As in previous work~\cite{saywell2022efficient}, we apply penalties to enforce waveform smoothness and constrain the peak Rabi frequency to below the nominal value $\Omega_0$. Restricting the peak Rabi frequency is critical to ensure a fair comparison between different pulses.

To obtain a beamsplitter pulse with variable (but robust) pulse origin, we use the following state-to-state fidelity measure:
\begin{equation}\label{eq:fidelity}
    \mathcal{F}(\delta)=\bigg|\frac{1}{\sqrt{2}}(\bra{g}+e^{-i\Phi(\delta)}\bra{e} )\hat{U}\ket{g}\bigg|^2.
\end{equation}
Here, $\hat{U}$ represents the shaped pulse propagator (defined by Equation~\eqref{eq:prop-shaped}) and $\Phi(\delta)$ is the phase dispersion, which we choose to be linear in $\delta$ (meaning $\Phi=m\delta$) and which directly relates to the pulse origin through Equation~\eqref{eq:bs-origin}. The second beamsplitter pulse is obtained using the ``flip-reverse'' procedure introduced in~\cite{saywell2020optimal}. This ensures that the phase dispersions of the first and final pulses cancel, keeping the interferometer insensitive to atomic velocity. 

To design pulses with stable temporal origins, we average the fidelity over a 2-dimensional ensemble of detuning $\delta$ and pulse amplitude error $\epsilon$ (defined via ${\tilde{\Omega}(t)\rightarrow(1+\epsilon)\tilde{\Omega}(t)}$) at each iteration of the optimization. By averaging $\mathcal{F}$ over a range of $\epsilon$ (taken to be $\pm 10\%$), we engineer robustness of the phase dispersion (and hence pulse origin) to fluctuations in laser intensity. Similarly, we average $\mathcal{F}$ over a range of detunings (in the range $\pm 1.5\Omega_0$) to enhance velocity acceptance and ensure $\tau^o$ remains constant as the detuning is varied. 

Designing control pulses with tailored phase dispersions has previously been explored in the context of nuclear-magnetic-resonance (NMR). Here, pulses designed with linear~\cite{gershenzon2008linear, koos2015broadband, koos2017broadband2, coote2021optimization} and quadratic~\cite{goodwin2020second} phase dispersion have both been shown to increase bandwidth for a given pulse duration. In particular, the Inherent Coherence Evolution optimized Broadband Excitation Resulting in constant phase Gradients (ICEBERG) pulses of Ref.~\cite{gershenzon2008linear} are analogous to the beamsplitter pulses with variable pulse origin we design in this work for robust interferometric sensing.

Figure~\ref{fig:optimised_pulse_comparison} compares three different beamsplitter pulses: the rectangular $\pi/2$ pulse, a ``point-to-point" pulse designed with the origin coincident with the pulse endpoint (taken from Ref.~\cite{saywell2021can}), and an ``optimized origin" pulse designed with its origin within the pulse using Equation~\eqref{eq:fidelity}. Figures~\ref{fig:optimised_pulse_comparison} (a) and (d) show how the phase dispersion and final excited state probability following each pulse depend on the detuning, respectively. The pulse with its origin within the pulse has a better frequency (velocity) acceptance than both the rectangular beamsplitter and the point-to-point pulse from Ref.~\cite{saywell2021can}. It is also less than half as long as the pulse from Ref.~\cite{saywell2021can}. 

Figures~\ref{fig:optimised_pulse_comparison} (b) and (e) illustrate how these two optimized beamsplitters differ in their respective temporal origins. This is shown by extrapolating the phases accumulated for a range of differently detuned atoms back to a common point in time. Figures~\ref{fig:optimised_pulse_comparison} (c) and (f) show how the different pulse origins relate to the trajectories followed by atoms on the Bloch sphere: the point-to-point pulse refocuses the atomic spins to a single point on the equator whereas the optimized temporal origin pulse directs them to a line of constant latitude. 

\begin{figure}[!tb]
\includegraphics[width=\columnwidth]{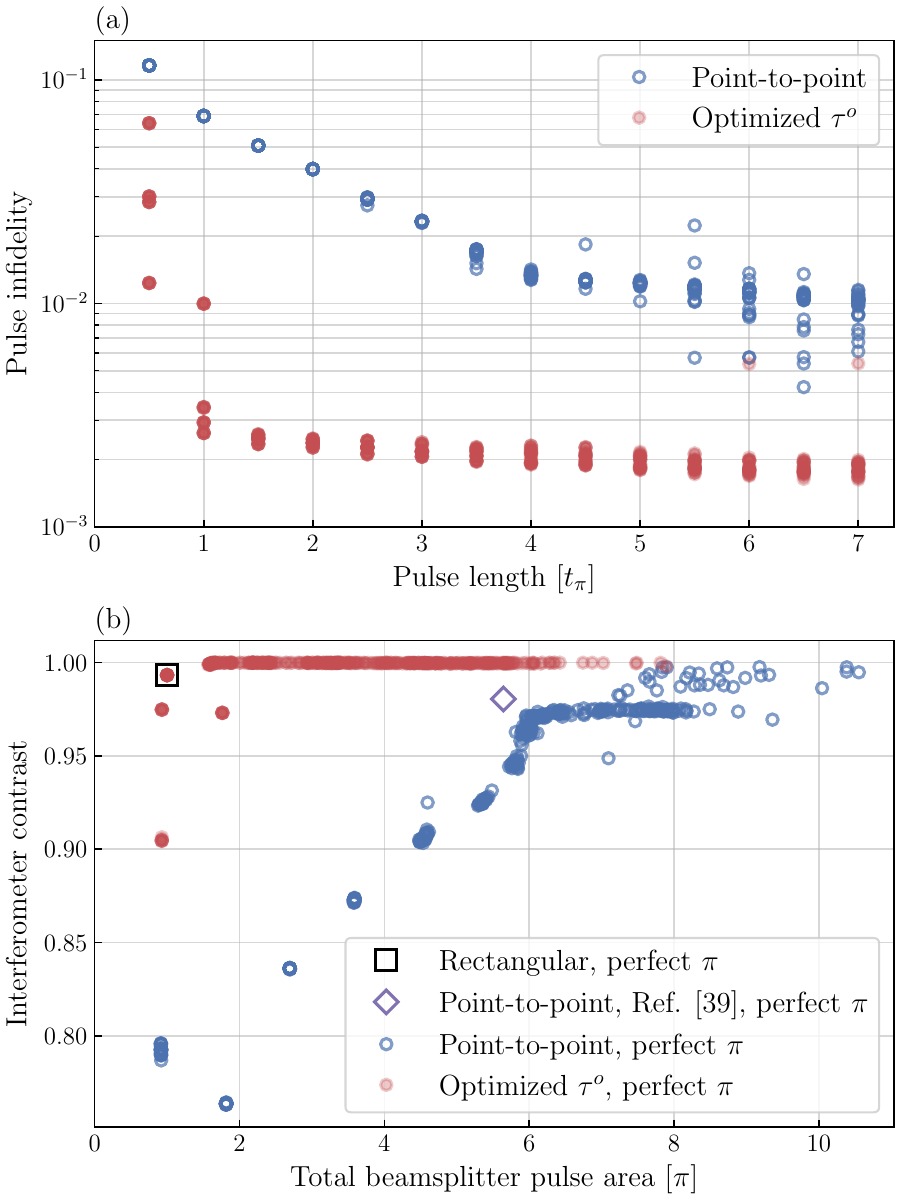}
\caption{\label{fig:batch_comparison} (a) Terminal infidelity of optimized beamsplitters as the pulse duration is varied (shown as multiples of the duration of a $\pi$ pulse). 28 optimizations of each beamsplitter class (point-to-point and optimized $\tau^o$) were performed for each choice of $\tau$. (b) Interferometer contrast for sequences composed of different beamsplitter pulses and the total beamsplitter pulse area. For each sequence, we assumed the mirror was ideal with zero Doppler sensitivity and that the atomic momentum distribution was Gaussian with $1\sigma=0.4\ \hbar k$. The blue empty circles correspond to point-to-point pulses where the origin is at the end of the pulse ($\tau^o=\tau$). The red filled circles correspond to optimized origin pulses which are designed such that the origin lies within the pulse ($\tau^o<\tau$). The black square corresponds to a sequence of rectangular beamsplitters and the purple diamond a sequence formed using the point-to-point beamsplitter from~\cite{saywell2021can}. The spread in points in (a) and (b) is due to the fact that a different random guess was used to optimize each pulse.
}
\end{figure}

Designing beamsplitters with stable origins within the pulse proves to be a less restrictive optimization problem than minimizing the gradient of the phase dispersion entirely, allowing us to find shorter pulses of equivalent fidelity. To demonstrate this, we optimized a selection of beamsplitters with different pulse origins and durations starting from initial random waveforms. Specifically, we varied the pulse duration between $t_\pi/2$ to $7t_{\pi}$ in steps of $t_\pi$/2 with 50 time-slices per $\pi$ pulse duration $t_\pi$. For each duration, we performed 28 optimizations for $\tau_o=\tau$ (point-to-point pulses) and 28 for $\tau_o<\tau$ (7 per origin choice $0.2\tau$, $0.4\tau$, $0.6\tau$, and $0.8\tau$). The infidelities obtained after 400 iterations for each duration are shown in Figure~\ref{fig:batch_comparison} (a) for each class of pulse: point-to-point and optimized $\tau^o$. We find that beamsplitters with temporal origins within the pulse reach a consistently lower infidelity as the pulse duration (and hence maximum pulse area) is shortened. 

Lower pulse infidelity should translate to a higher interference fringe contrast. Figure~\ref{fig:batch_comparison} (b) shows how the interferometer contrast relates to the total beamsplitter pulse area for a selection of sequences composed of beamsplitters with different pulse origins. We have also included the point-to-point pulse from~\cite{saywell2021can} as a point of comparison. We calculate the interference contrast numerically using the approach presented in~\cite{gillot2016limits} assuming a $\pi/2$ pulse duration of 10~$\mu$s and two-photon Raman transitions performed using counter-propagating beams on the D2 line in $^{85}$Rb. The atomic momentum distribution is assumed to be Gaussian with $1\sigma=0.4\ \hbar k$, where $k$ is the effective wave-number for the Raman transition~\footnote{Choosing to simulate an atomic momentum distribution with a $6\sigma$ width of $3\Omega_0$ in detuning means that 99.7\% of atoms will be within the optimization ensemble.}.

We further assume that the mirror pulse is a ``perfect'' $\pi$ pulse meaning it has zero Doppler sensitivity. This allows us to isolate the effect of the beamsplitters on the fringe contrast for a given momentum distribution. When the peak Rabi frequencies of all rectangular pulses in the interferometer are identical, it is the $\pi$ pulse that most limits the interferometer's velocity acceptance~\cite{saywell2018optimal} and therefore the interference fringe contrast. 

\begin{figure*}[!t]
\includegraphics[width=\linewidth]{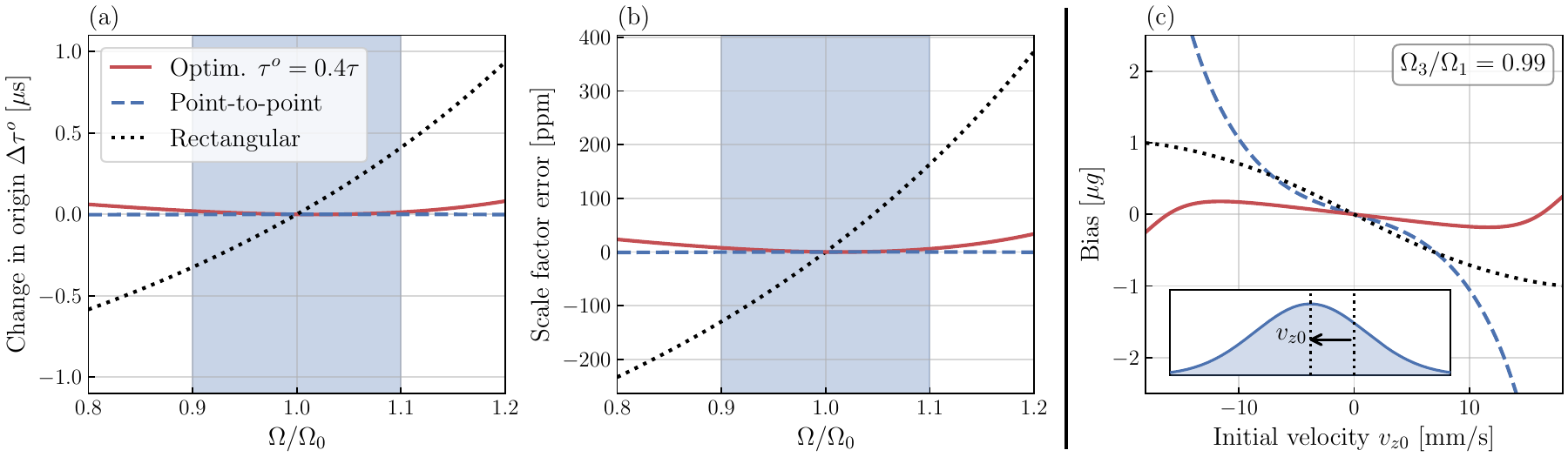}
\caption{\label{fig:scale_factor_stability}(a) The change in temporal origin of three different beamsplitter pulses as the peak Rabi frequency is varied as a fraction of the nominal peak value $\Omega_0$. (b) Scale factor error in parts per million calculated for Mach-Zehnder sequences (assuming $T=5$~ms) composed of the same three beamsplitter pulses as the peak Rabi frequency of all pulses is varied. The blue shaded region corresponds to the optimization ensemble range for amplitude robustness. (c) Simulated accelerometer bias in $\mu g$ for an interferometer with $T=5$~ms using the same beamsplitter pulses depicted in (a, b) as the mean initial atomic velocity is varied and the Rabi frequency of the final pulse is 1\% lower than the first. The Gaussian atomic momentum distribution with $1\sigma=0.4\ \hbar k$ is shown in the inset.}
\end{figure*}

We find that for this momentum width the point-to-point pulses consistently yield marginally lower fringe contrasts than the $\pi/2$ pulses at shorter durations. This is a direct consequence of their lower individual fidelity. With the optimized origin pulses, however, we are able to maintain and even improve fringe contrast with less than $\pi$ radians of additional area. For a given duration, the $\tau^o<\tau$ pulses use less of the available maximum pulse area to reach a given fidelity and we find that for each duration on average the areas for these pulses are $1.5\times$ lower than the $\tau^0=\tau$ pulses. A reduction in pulse area means a reduction in spontaneous emission. In interferometers that use two-photon Raman transitions this means the interferometer can be operated closer to single-photon resonance (thereby increasing the two-photon Rabi frequency and sequence velocity acceptance) for a given optical power.

\subsection{Robustness to laser intensity fluctuations}

For interferometers composed of conventionally-shaped pulses, any variation in the laser intensity (and therefore Rabi frequency) ``seen'' by the atoms changes the temporal origins of the $\pi/2$ pulses and hence the interferometer's scale factor. If the laser intensity varies temporally \textit{between} the $\pi/2$ pulses, then the symmetry of the interferometer breaks~\cite{gillot2016limits}, making it sensitive to fluctuations in the initial velocity of the atoms. Both these effects are detrimental in precision inertial and gravitational sensing. 

Variations in laser intensity naturally arise due to fluctuations and drifts in laser power. Pulse-to-pulse variations in laser intensity during a sequence arise due to thermal expansion of the atomic cloud across non-uniform (e.g. Gaussian) interferometry beams. They can also be caused by platform accelerations directed transverse to the optical propagation (measurement) axis which displace the atomic cloud to a region of lower intensity in the beams~\cite{saywell2023enhancing, dedes2023optimizing}. Fluctuations in initial atomic velocity (or, more precisely, mean initial atomic detuning) can be caused by motion of the platform parallel to the measurement axis, magnetic field gradients, and variations in atom trap release time~\cite{gillot2016limits}.

\subsubsection{Scale factor stability}

We confirm that minimizing variation in the pulse origin with laser intensity enhances scale factor stability by calculating both quantities for a range of Rabi frequencies centered on a nominal value of $\Omega_0/2\pi=25$~kHz. The results are shown in Figures~\ref{fig:scale_factor_stability} (a) and (b) for three different beamsplitter pulses: the rectangular $\pi/2$ pulse, an optimized point-to-point pulse designed such that $\tau^o=\tau$ (origin at the end of the pulse), and another optimized origin pulse with $\tau^o =0.4\tau$ (origin within the pulse). The two optimized beamsplitters in Figure~\ref{fig:scale_factor_stability} correspond to those with the least variation in $\tau^o$ over the design range of amplitude errors ($\pm 10\%$). As expected from Equation~\eqref{eq:rectangular-origin}, the rectangular beamsplitter's temporal origin changes appreciably (by 735~$\mathrm{n}$s) as the Rabi frequency is varied from 0.9 to $1.1\Omega_0$. Conversely, we see that the pulse origins remain stable for both the point-to-point and optimized $\tau^o$ pulses, varying by 1 and 21~$\mathrm{n}$s over the same amplitude range, respectively. 

The enhancement in pulse origin stability directly translates to an improved scale factor stability in the presence of laser intensity fluctuations. This is shown in Figure~\ref{fig:scale_factor_stability} (b) where we have computed the interferometer scale factor by numerically integrating the sensitivity function for sequences constructed using the same beamsplitters used in (a). A rectangular mirror pulse was used for each sequence. The interferometer's interrogation time was 5~ms, which is comparable to that used in mobile interferometers~\cite{geiger2011detecting, barrett2016dual}. The maximum scale factor errors for the point-to-point and optimized $\tau^o$ pulses are 0.4~ppm and 7.7~ppm, respectively, as the Rabi frequency of the final pulse is varied by $\pm10\%$ of the first. In contrast, the maximum scale factor error for rectangular pulses is 163.6~ppm. For this particular optimized $\tau^o$ pulse, the reduction in scale factor error is achieved with just $\pi$ radians of additional pulse area compared with $2.5\pi$ for the point-to-point beamsplitter.

We can express the fractional scale factor error in terms of the change in temporal origins $\Delta\tau_i^o$ as follows:
\begin{equation}\label{eq:fractional-scale-factor-error}
    \mathcal{S_{\epsilon}}\approx \frac{1}{T}(\Delta\tau_3^o-\Delta\tau_1^o).
\end{equation}
The scale factor error caused by fluctuating pulse origins therefore decreases as the ratio $T/\tau$ increases. State-of-the-art mobile interferometers operating on moving platforms are currently limited in their maximum interrogation times to $\sim 20$~ms~\cite{bidel2018absolute, bidel2020absoluteairborne, barrett2016dual, geiger2011detecting} by a combination of platform rotations~\cite{barrett2016dual} and vibrations~\cite{bidel2020absoluteairborne}. Even at $T=20$~ms, however, the maximum scale factor error caused by $\pm10\%$ fluctuations in laser intensity remains at 41~ppm for rectangular pulses operating with a typical nominal Rabi frequency of $\Omega_0/2\pi=25$~kHz. This is reduced to just 2~ppm for the $\tau^0<\tau$ optimized origin beamsplitter.

\subsubsection{Interferometer symmetry}\label{sec:symmetry}

Improving pulse origin stability against temporal intensity fluctuations also implies that the interferometer will remain closed when there is an imbalance in Rabi frequencies during the sequence. This reduces the bias caused by a non-zero initial atomic detuning and hence non-zero initial atomic velocity. This remains a problem even in interferometers where a narrow velocity slice is selected from the initial atomic distribution by using, for example, a long, frequency-selective $\pi$ pulse immediately prior to the interferometer sequence. In this case, any variation between the selection pulse frequency and the initial atomic velocity results in a skewed selected velocity distribution with a center~\cite{gillot2016limits} shifted relative to the selection frequency.

Close to resonance (and in the limit of infinitely narrow momentum width) the phase sensitivity to variations in initial mean atomic velocity $v_z$ is given by
\begin{equation}\label{eq:generalized-phase-sensitivity}
    \frac{\partial\Phi}{\partial v_z}=k(\Delta\tau_1^o+\Delta\tau_3^o-2\Delta\tau_2^o),
\end{equation}
where $\Delta \tau_i^o$ is the change in pulse origin induced by a fluctuation in e.g. Rabi frequency. The sensitivity to non-zero mean initial atomic velocity is illustrated in Figure~\ref{fig:scale_factor_stability} (c) where we show the simulated bias for a 5~ms interferometer calculated using different choices of beamsplitters as the mean initial velocity of the atomic cloud is varied. The phase here is averaged over a Gaussian atomic momentum distribution with $1\sigma=0.4\ \hbar k$ for an imbalance in Rabi frequencies of 1\% between the first and  final beamsplitters. Such an imbalance could feasibly arise in a mobile setting and could be caused, for example, by a constant $\sim$~0.7~$g$ platform acceleration directed perpendicular to the measurement axis $z$ if using Gaussian beams with a typical $1/e^2$ diameter of 10~mm.

This small intensity difference is sufficient to result in a non-zero mean velocity sensitivity $|\partial \Phi/\partial v_{z0}|$ of 0.32~mrad/(mm/s) for rectangular pulses near resonance. This is reduced to 0.07 and 0.19~mrad/(mm/s) by the optimized origin and  point-to-point sequences, respectively, for velocities close to resonance in the range -2 to 2~mm/s. For rectangular sequences, the increased velocity sensitivity yields a $709 \ \mathrm{n}g$ accelerometer bias $b_z$ when the initial velocity is offset from resonance by 10~mm/s. Using the optimized beamsplitter pulse with $\tau^o<\tau$ reduces this bias to 172~$\mathrm{n}g$, which equates to a 4$\times$ reduction.

Although the bias for the point-to-point beamsplitter is less than that for the rectangular pulses close to resonance, it becomes far more sensitive as the mean atomic velocity offset increases. We attribute this fact to the lower fidelity reached by this pulse in the optimization compared with the optimized origin pulse (98.7\% vs. 99.8\%). The lower fidelity of this pulse means that, for a given atom, its phase gradient with respect to detuning is more sensitive to variations in Rabi frequency further from resonance. This therefore increases the phase sensitivity of this beamsplitter to small fluctuations in initial mean velocity when averaging over a broad momentum distribution. For example, if we reduce the width of the momentum distribution such that $1\sigma=0.05\ \hbar k$, then the bias sensitivities for the optimized origin and point-to-point pulses are reduced to just 0.02~mrad/(mm/s) close to resonance, while the rectangular pulse's bias sensitivity increases to 0.57~mrad/(mm/s).

\section{Systematic errors in Doppler compensation}

\begin{figure}[tb]
\includegraphics[width=\columnwidth]{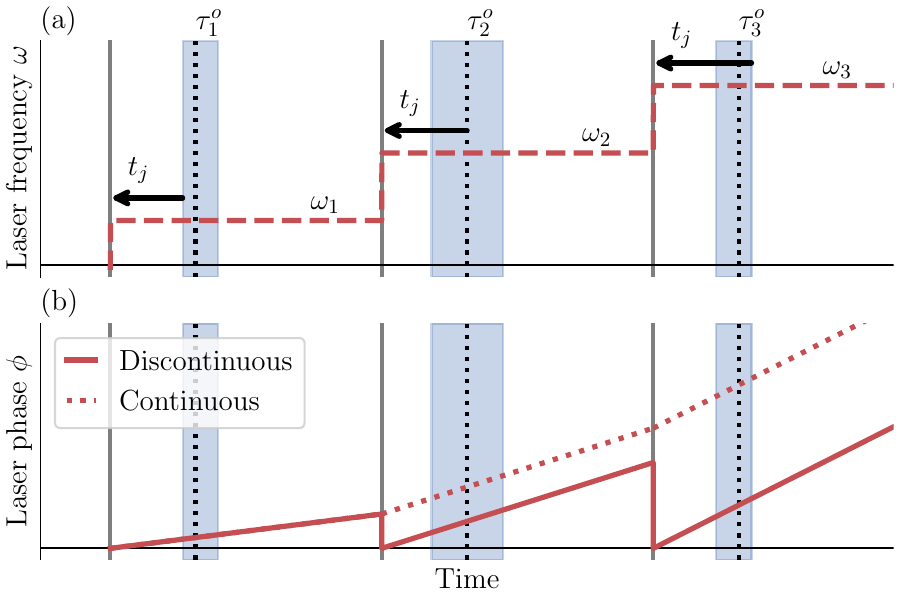}
\caption{\label{fig:doppler-compensation} 
Doppler compensation schemes using laser frequency jumps to maintain resonance in a Mach-Zehnder interferometer~\cite{templier2022tracking}. Laser frequency (a) and phase (b) are shown for phase-discontinuous and phase-continuous cases. In the phase-discontinuous case, the laser phase is reset whenever the frequency is changed. This cancels the phase accumulated by the atom if the frequency jumps all occur at a fixed time prior to the temporal origin of each pulse. We have adopted the timing convention of Ref.~\cite{templier2022tracking} where frequency jumps occur at a time $t_j$ prior to the beginning, middle, and end of pulses 1, 2, and 3, respectively.
}
\end{figure}

To keep the interferometry laser on resonance throughout a pulse sequence, its frequency must be changed to compensate for the time-dependent Doppler shift caused by atomic acceleration in the rest frame of the laser. The details of the experimental implementation of this frequency change matter because the interferometer phase is given by the difference between the phase accumulated due to atomic motion and the phase imprinted by the laser. In cases where the acceleration is constant during the pulse sequence (such as a lab-based gravimeter), the Doppler detuning varies linearly with time. In these cases, applying a continuous frequency chirp to the laser frequency such that $\dot{\omega}=\alpha$ fully compensates the Doppler shift if $\alpha=ka$. In the limit where $T\gg\tau$, the total phase of the interferometer is given by $k(a-\alpha)T^2$, meaning the interferometer can be locked, for example, to a central fringe using feedback to the laser chirp rate to maintain a zero phase output~\cite{louchet2011influence}.

In many experiments, the laser phase and frequency are determined using a Direct Digital Synthesizer (DDS)~\cite{templier2022tracking, peters1999measurement}. In these cases, a continuous frequency chirp can be emulated using a series of small discrete frequency steps. On moving platforms, however, the acceleration may change appreciably from shot-to-shot and even pulse-to-pulse. In these cases, it is more convenient to make a single adjustment in frequency before each pulse to ensure that pulse is on resonance. These jumps in frequency may be continuous or discontinuous in phase, as shown in Figure~\ref{fig:doppler-compensation}.

However, differences between the laser and inertial phases accrued by the atoms during the interferometer can result in systematic phase shifts~\cite{templier2021three, templier2022tracking, templier2021carrier}. These phase shifts - which were listed as the dominant contribution to the error budget in the 3-axis hybrid quantum accelerometer of Ref.~\cite{templier2022tracking} - bias the measurement and reduce stability. In this section, we explain how the temporal pulse origin provides a straightforward explanation of these systematic errors and show how they may be eliminated by using pulses with stable origins and appropriate timing of frequency jumps.

\subsection{Phase-continuous frequency jumps}

Let’s first consider the case where Doppler compensation is performed using phase-continuous frequency steps between the pulses. Suppose, following~\cite{templier2022tracking}, that the laser frequency is stepped at a time $t_j$ prior to the beginning, middle, and end of pulses 1, 2, and 3, respectively. The total laser phase $\Phi_L$ accumulated by the atoms is given by
\begin{equation}\label{eq:laser-phase-continuous}
    \Phi_L=\int_{-\infty}^{+\infty}g(t)\delta_L(t)\,\mathrm{d}t.
\end{equation}
$\delta_L(t)=\omega_{eg}-\omega(t)$ is the laser detuning, which takes the value $\delta_{Li}$ during pulse $i$. In order to compensate the Doppler shift due to an acceleration $a$, we require
\begin{equation}
    \delta_{L2}-\delta_{L1}=\delta_{L3}-\delta_{L2}=-ka(T+\tau+\tau_2/2),
\end{equation}
where, for simplicity, we have assumed that the durations of the first and final beamsplitter pulses are identical and that the mirror pulse is rectangular (meaning its origin is at the pulse center). Recalling that the area under $g(t)$ during pulse $i$ is approximately given by the gradient of its phase dispersion $m_i$, we find
\begin{align} \nonumber
    \Phi_L=&\sum_i\delta_{Li}m_i+(\delta_{L2}-\delta_{L1})(T+\tau_2/2-t_j)\\&+(\delta_{L3}-\delta_{L2})(t_j-\tau).
\end{align}
The above expression simplifies to
\begin{equation}
    \Phi_L\approx \delta_{L1}(m_1+m_3)-ak(T^2+T\tau_2+2m_3T),
\end{equation}
where we have dropped terms which are second-order in pulse durations. Note that in the phase-continuous scheme the total laser phase does not depend on the precise time $t_j$ at which the frequency jumps occur. 

The inertial phase can be found using the scale factor derived in Section~\ref{sec:shaped-pulses}
\begin{equation}
    \Phi_I\approx kv_0(m_1+m_3)+ak(T^2+T\tau_2-m_1T+m_3T).
\end{equation}
Here, $kv_0$ is the Doppler detuning during the central $\pi$ pulse. If the change in detuning exactly matches the inertial acceleration, the total interferometer phase $\Phi=\Phi_L+\Phi_I$ becomes
\begin{equation}\label{eq:tot_phase_cont}
    \Phi=(kv_0+\delta_{L1}-akT)(m_1+m_3).
\end{equation}
At first inspection of the acceleration-dependent phase it appears that a scale factor error is introduced by a discrepancy between the laser and inertial phases. However, the frequency $k v_0 + \delta_{L1} - a k T$ in Eq.~\eqref{eq:tot_phase_cont} is approximately equal to the total detuning during the mirror pulse. In other words, the systematic error reported in Ref.~\cite{templier2022tracking} for Doppler compensation implemented using phase-continuous frequency jumps can be understood as the sensitivity to a constant detuning from resonance caused by a break in the symmetry of the interferometer. As explained in Section~\ref{sec:symmetry}, this error may be suppressed by using beamsplitter pulse pairs designed with stable origins.

\subsection{Phase-discontinuous frequency jumps}

\begin{figure}[tb]
\includegraphics[width=\columnwidth]{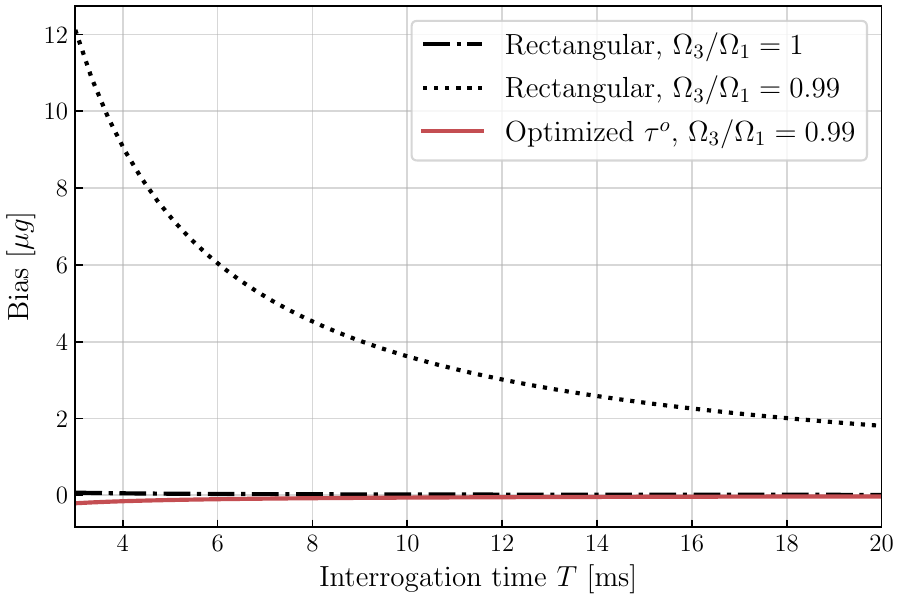}
\caption{\label{fig:discontinuous-phase-error} 
Interferometer bias caused by phase-discontinuous Doppler compensation using rectangular and optimized origin beamsplitters as the interrogation time is varied. When $\Omega_1\neq \Omega_3$, the timing adjustment $\Delta t_j\approx \tau-2\tau/\pi$ no longer completely nullifies the laser phase for rectangular pulses because the pulse origins are shifted by the Rabi frequency imbalance. However, the corresponding timing adjustment remains valid for the optimized origin beamsplitter used in Fig.~\ref{fig:scale_factor_stability} and the bias phase is therefore minimized in this case even when the Rabi frequencies are imbalanced. 
}
\end{figure}

Now we consider the case where Doppler compensation is performed using phase-discontinuous frequency jumps. Here, the running phase accumulated by the laser field is subtracted or “reset” whenever the frequency is jumped. The objective of this scheme is to completely nullify $\Phi_L$, ensuring that the interferometer’s phase depends solely upon atomic acceleration and not the laser frequency difference between pulses. We will show that in order for $\Phi_L$ to completely vanish, the frequency jumps should be timed such that they all occur at a fixed time prior to the origin of each pulse. Anticipating this, we therefore adjust the timings of the frequency jumps for pulses 1 and 3 as follows:
\begin{align}\label{eq:timing-adjustment1}
    t_{j1}&=t_j-\Delta t_j, \\ \label{eq:timing-adjustment2}
    t_{j3}&=t_j+\Delta t_j.
\end{align}
Now, the total laser phase is given by
\begin{equation}\label{eq:laser-phase-discontinuous}
\Phi_L=\int_{-\infty}^{+\infty}g(t)\delta_L(t)\,\mathrm{d}t+\phi_{1\rightarrow 2}-\phi_{2\rightarrow 3},
\end{equation}
where $\phi_{1\rightarrow 2}=\delta_{L1}(\tau+T+\tau_2/2-\Delta t_j)$ and $\phi_{2\rightarrow 3}=\delta_{L2}(\tau+T+\tau_2/2-\Delta t_j)$ are the phases that are subtracted from the laser during a frequency step. Equation~\eqref{eq:laser-phase-discontinuous} simplifies to
\begin{equation}\label{eq:laser-phase-discontinuous-simple}
    \Phi_L\approx(m_1+m_3)\delta_{L2}-akT(m_3-m_1+2\Delta t_j-2\tau),
\end{equation}
where we have again dropped terms that are second-order in pulse durations. In the case where $m_1=-m_3$ (e.g. closed interferometer formed from rectangular pulses with constant Rabi frequency), we see that the laser phase vanishes completely if we choose $\Delta t_j$ such that all frequency jumps occur at a fixed time prior to the origins of each pulse: $\Delta t_j=\tau-m_3$. For rectangular pulses with ideal pulse area, we find $\Delta t_j\approx \tau-2\tau/\pi$, in agreement with~\cite{templier2022tracking}.

The reason the timings must be adjusted in this way is because the laser accumulates phase at a constant rate while the atom only accumulates phase at a constant rate during the interrogation times, causing a mismatch between the two. In Equation~\eqref{eq:laser-phase-discontinuous} we subtract the phase $\phi_d=\phi_{2\rightarrow 3}-\phi_{1\rightarrow 2}$ from the phase accumulated by the atoms due to the frequency jumps. The phase $\phi_d$ is in fact equivalent to the phase of an ideal Mach-Zehnder atom interferometer operating with infinitesimal pulses separated by $T_{\mathrm{eff}}=\tau+T+\tau_2/2-\Delta t_j$; $\phi_d$ is simply the difference in phases accumulated in the two periods of free evolution. Therefore for this phase to cancel with that accumulated by the atoms in the real interferometer with finite pulses (Equation~\eqref{eq:laser-phase-continuous}), the effective interrogation time $T_{\mathrm{eff}}$ must match the real interrogation time, which is given by the time between pulse origins: $T_{\mathrm{eff}}=\tau_2/2+T+m_3$. This is satisfied if $\Delta t_j=\tau-m_3$.

We see from Equation~\eqref{eq:laser-phase-discontinuous-simple} that if the pulse origins for the beamsplitters vary (e.g. due to a laser intensity imbalance between pulses 1 and 3) then the timing adjustment required to nullify the phase will also change. Unless $\Delta t_j$ can be updated accordingly, this will result in a non-zero bias. However, the use of beamsplitters with stable temporal origins will ensure that a single timing adjustment is sufficient to nullify the laser phase even in the presence of laser intensity fluctuations. This is shown in Figure~\ref{fig:discontinuous-phase-error}, where we have numerically calculated the total laser phase $\Phi_L$ for sequences composed of rectangular beamsplitters and the optimized origin beamsplitters used in Figure~\ref{fig:scale_factor_stability}. In the case of a 1\% Rabi frequency imbalance between pulses 1 and 3, the optimized origin beamsplitters reduce the magnitude of the residual phase bias caused by phase-discontinuous frequency jumps from 7.21 to 0.16~$\mu g$ at $T=5$~ms.

\section{Discussion}

The inertial phase response of any finite-duration pulse in an interferometric sensor may be characterized by a single point in time, which we refer to as the pulse's temporal origin. We have shown how knowledge of the temporal origins of pulses in an interferometer sequence permits a simple but accurate determination of its measurement scale factor. A stable pulse origin is therefore critical to maintain scale factor stability and minimize bias in the presence of experimental inhomogeneities such as temporal fluctuations in laser intensity. Stable pulse origins also preserve the interferometer's temporal symmetry - which is required in spin-echo type sequences (e.g. accelerometers and gyroscopes) for the interferometer to remain insensitive to atomic velocity. 

We also note that the pulse origins are highly relevant when correcting for the effect of platform vibrations in real-time. Here, the signal from a mechanical accelerometer is used to estimate and subtract a vibrational phase from the atom interferometer~\cite{templier2022tracking}. The triangular acceleration response function is used to calculate these corrections. The pulse origins are the optimal times at which to place the vertices of the response function in order to maximize correlations between the classical and quantum sensors.

We have also explored how the pulse origin can be incorporated as a design parameter in the optimization of robust pulses. Previous approaches have mainly relied upon designing pulses using so-called ``point-to-point'' or ``universal rotation'' fidelity measures originally developed for NMR spin control. For beamsplitter and mirror pulses, these design objectives keep the temporal origins fixed at the end or center of the pulses, respectively. We have shown that for individual beamsplitters, we can achieve better fidelity with less pulse area by relaxing this constraint and placing the origin within the pulse. This has several advantages. For example, in interferometers using two-photon Raman transitions, reducing the pulse area lowers the spontaneous emission, allowing us to make the most of available optical power by operating closer to single-photon resonance. Moreover, the interferometer's scale factor remains calculable and stable in the presence of laser intensity variations.

The figure of merit used to design optimized beamsplitters in this work equally weights population transfer and phase dispersion. Consequently, we found that some pulses that reached high terminal fidelities in the optimization had low phase gradient stability against Rabi frequency fluctuations and vice versa. This could be improved, for example, by preferentially weighting phase dispersion over population transfer in the fidelity~\cite{skinner2005tailoring}. Alternatively, one could incorporate the interferometer's sensitivity or response functions into tailored figures of merit by adapting the perturbative approach introduced in~\cite{dedes2023optimizing}.

In future work, we will apply the pulse origin approach to the design of mirror pulses and alternative interferometer configurations such as Ramsey and large-momentum-transfer sequences. More generally, the pulse origin allows us to treat the design of entire sequences as a single optimization problem (namely, the temporal placement and stability of pulse origins) rather than a series of individual pulse optimizations. Finally, we anticipate that the pulse origin may be applied to any quantum sensor reliant upon sequences of finite-duration control pulses.

\begin{acknowledgments}
The NMR spin dynamics simulation software Spinach~\cite{hogben2011spinach} and its optimal control module were used to design the shaped pulses in this work. The Python package QuTiP~\cite{johansson2012qutip} was used to produce Bloch sphere diagrams. The authors are grateful for funding from the UK Engineering and Physical Sciences Research Council under grant No. EP/Y035267/1.
\end{acknowledgments}

\appendix

\section{Atom-light interactions in interferometry}\label{appendix:atom-light}

The Hamiltonian for a two-level atom with state vector $\ket{\psi(t)}=c_g(t)\ket{g}+c_e(t)\ket{e}$ interacting with a monochromatic optical field with frequency $\omega$, phase $\phi$ and wave-vector $\mathbf{k}$ can be written in the rotating frame as $\mathbf{\Omega}\cdot\hat{\mathbf{\sigma}}/2$. $\hat{\mathbf{\sigma}}$ is the vector of Pauli matrices and $\mathbf{\Omega}$ is the field vector:
\begin{equation}
\mathbf{\Omega}=\Omega\cos\phi\hat{\mathbf{x}}+\Omega\sin\phi\hat{\mathbf{y}} - \delta\hat{\mathbf{z}}.
\end{equation}
$\Omega$ is the Rabi frequency, and $\delta=\omega_{eg}-\omega+\mathbf{k}\cdot \mathbf{v}$ is the detuning of the driving field from the transition $\omega_{eg}$. The term $\mathbf{k}\cdot \mathbf{v}$ corresponds to the Doppler shift for an atom with initial velocity $\mathbf{v}$. 

For interferometers which employ two-photon Raman transitions, the same equations govern pulse evolution with $\Omega$, $\mathbf{k}$, $\omega$, and $\phi$ replaced by their two-photon equivalents~\cite{kasevich1991atomic}.

The temporal evolution $\ket{\psi(t)}\rightarrow\ket{\psi(t+\tau)}$ under a rectangular pulse of duration $\tau$ with fixed Rabi frequency is given by application of the following propagator:
\begin{equation}\label{eq:prop}
    \hat{U}=\exp\big(-{i}\mathbf{\Omega}\cdot\hat{\mathbf{\sigma}}\tau/2 \big).
\end{equation}
In the Bloch sphere picture, we recognize this operation as a rotation of the state vector $\braket{\psi(t)|\hat{\mathbf{\sigma}}|\psi(t)}$ by an angle $|\mathbf{\Omega}|\tau$ about the normalized field vector.

As explained in the main text, in this work we consider shaped pulses described by piecewise-constant waveforms in generalized Rabi frequency $\tilde\Omega(t)=\{\Omega^n\cos\phi^n\}$. Here, $n=\{1,\ ...,\ M\}$ runs over the $M$ time-slices, which for convenience we restrict to the same duration $\Delta t$. The temporal evolution under a shaped pulse can be found by constructing a propagator given by the time-ordered products of propagators for each time-slice:
\begin{equation}\label{eq:prop-shaped}
    \hat{U}_{\mathrm{shaped}}=\hat{U}_M\hat{U}_{M-1}\ldots\hat{U}_2\hat{U}_1.
\end{equation}
The propagator for each time-slice (for which $\Omega$ and $\phi$ are held constant) is given by Equation~\eqref{eq:prop}. The action of a shaped pulse still corresponds to a rotation on the Bloch sphere, though the rotational axis, angle, and their dependence upon detuning and peak pulse amplitude are no longer immediately apparent.

\section{Rectangular mirror pulse as two back-to-back beamsplitters}\label{appendix:mirror-dead-time}

\begin{figure}[tb]
\includegraphics[width=\columnwidth]{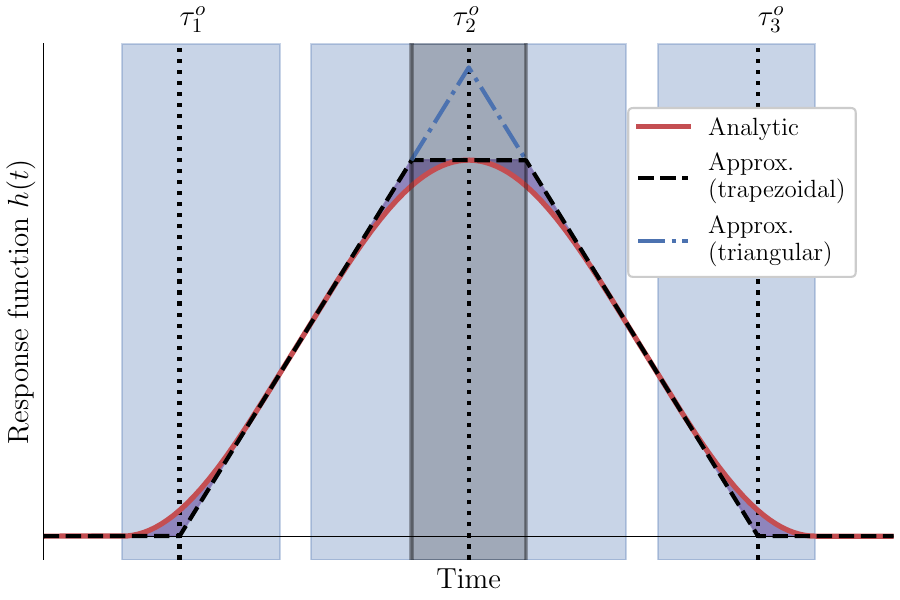}
\caption{\label{fig:rectangular-pulse-deadtime} 
 The inertial response function for a three-pulse Mach-Zehnder sequence composed of rectangular pulses. The interferometer scale factor is given by the area under the response function, which may be approximated as triangular in the limit where $T\gg\tau$. Considering the mirror as two back-to-back beamsplitters results in an improved trapezoidal approximation whose area exactly matches that given by the analytical result for $h(t)$. This is shown by the purple shaded regions, which represent the difference between the analytic and approximate response functions and which are exactly equal in area.
}
\end{figure}

The rectangular mirror pulse can be viewed as two back-to-back rectangular beamsplitter pulses. This gives rise to an effective ``dead-time'' during which no phase is accrued by a detuned atom, as indicated by the central shaded region in Figure~\ref{fig:rectangular-pulse-deadtime}. This dead-time is given by the separation between the origins of the two constituent beamsplitter pulses, hence ${\tau_d=2\tau-(2/\Omega)\tan(\Omega\tau/2)}$. Incorporating the effective dead-time of the mirror pulse yields an exact expression for the scale factor of a rectangular pulse interferometer. In this case the response function can be approximated as a trapezium whose top (upper) base is given by $\tau_d$:
\begin{align} \nonumber
    \mathcal{S} 
    =&\ \frac{k}{2} \bigg[
        2T + 2\tau 
        + \frac{2}{\Omega}\tan(\Omega\tau/2)
       + \tau_d
    \bigg] 
    \bigg[
        T + \frac{2}{\Omega}\tan(\Omega\tau/2)
    \bigg] \\ \label{eq:exact-rectangular-scale-factor}
   =&\ k(T + 2\tau)
    \bigg[
        T + \frac{2}{\Omega}\tan(\Omega\tau/2)
    \bigg].
\end{align}
This matches the result obtained by integrating the analytic expression for $h(t)$ (Equation~A4 in Reference~\cite{bonnin2015characterization}). Figure~\ref{fig:rectangular-pulse-deadtime} shows this graphically. The region between the trapezoidal approximation and the exact expression for $h(t)$ is depicted by four shaded purple regions of equal area. We can interpret this trapezoidal approximation as equivalent to the phase given by a Ramsey-Bord\'e interferometer where four infinitesimal $\pi/2$ pulses are located at the vertices of the trapezoid. The first and last $\pi/2$ pulses occur at the origins of pulses $1$ and $3$; and the central $\pi/2$ pulses occur at the origins of the two back-to-back beamsplitters that constitute the mirror, separated by $\tau_d$. 

Any mirror pulse will take a finite amount of time to transform the sensitivity function from $-1$ to $+1$. The area under the approximate triangular response function (which assumes instantaneous reflection at the pulse origin) will therefore always produce an over-estimate of the actual contribution of the mirror to the scale factor. This is accounted for in the rectangular case by defining a dead-time, leading to an improved (trapezoidal) approximation for the response function. In general, it is not possible to divide a shaped mirror pulse into two successive pulses that individually function as beamsplitters and a full generalization of this result to arbitrary sequences is left for future work. Nonetheless, it is still possible to define a dead-time for a shaped pulse sequence such that the exact scale factor is given by the area under a trapezoidal approximation to the response function~\footnote{This definition is valid when the area under $h^{\triangle}(t)$ for the entire sequence produces an over-estimate of the true scale factor.}:
\begin{equation}
    \tau_d^2=4\int_{-\infty}^{+\infty}[h^{\triangle}(t)-h(t)]\,\mathrm{d}t.
\end{equation}
$h^{\triangle}(t)$ is the triangular approximation to the true response function $h(t)$. The mirror pulse origin lies at the center of the upper base (of length $\tau_d$) of the trapezium.

\section{Scale factor calculation in open interferometers}\label{appendix:open-scale-factor}

\begin{figure}[tb]
\includegraphics[width=\columnwidth]{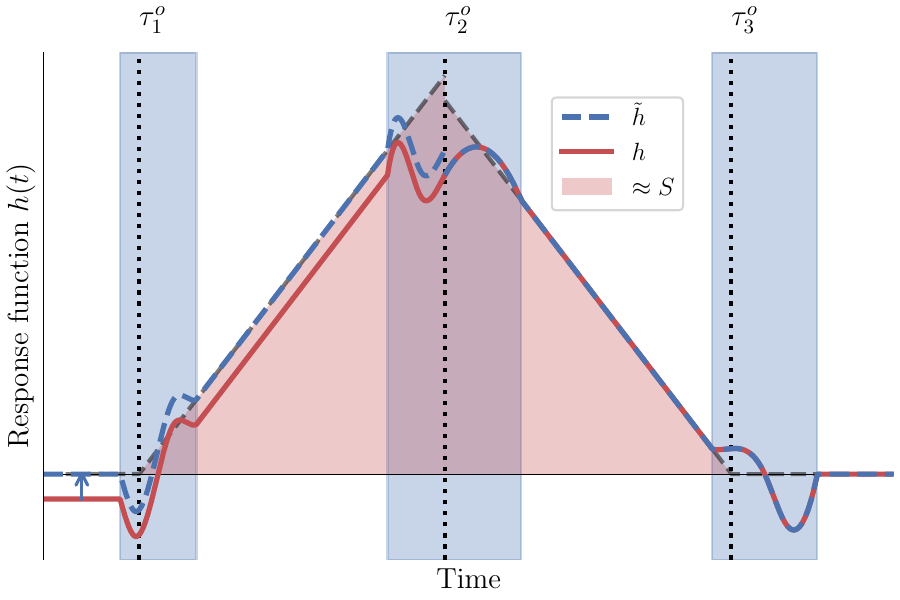}
\caption{\label{fig:pulse-origin-diagram-open} Original $h(t)$ and modified $\tilde{h}(t)$ response functions for an open interferometer composed of arbitrary shaped pulses. In the case where the interferometer is open, the response function is no longer zero at the beginning of the interferometer, indicating a sensitivity to atomic velocity $v_0$. By shifting the response function vertically by $h(t_i)$ during the first half of the sequence, we obtain the modified response function $\tilde{h}(t)$ whose area yields the approximate scale factor in the case where we define the velocity $v_0$ within the mirror pulse. The precise shape of the response function shown within each pulse is arbitrary for illustrative purposes.}
\end{figure}

In the general case of an open interferometer and constant acceleration $a$, the total interferometer phase can be decomposed into velocity-dependent and acceleration-dependent contributions:
\begin{align}
    \Phi &= \Phi_{v_0}+\Phi_a, \\
    &= \Phi_{v_0} + \mathcal{S}a.
\end{align}
$\Phi_{v_0}$ is the phase which depends upon atomic velocity $v_0$ defined at time $t_0$ and $\Phi_a=\mathcal{S}a$ is the acceleration phase proportional to the scale factor. Clearly, this separation of terms (and hence also the scale factor $\mathcal{S}$) depends upon where we define $t_0$. This matters when calculating the scale factor using the phase sensitivity and inertial response functions as we explain below.

In the presence of constant acceleration along $z$, the atomic detuning is given by
\begin{equation}
    \delta(t)=kv_0+ka(t-t_0).
\end{equation}
Using the phase sensitivity function $g(t)$, we can find the total interferometer phase as follows:
\begin{align}
    \Phi/k &= \int_{-\infty}^{+\infty}g(t)\big[v_0+a(t-t_0)\big]\,\mathrm{d}t, \\
    &= v_0\int_{-\infty}^{+\infty}g(t)\,\mathrm{d}t + a{\int_{-\infty}^{+\infty}g(t)(t-t_0)\,\mathrm{d}t}.
\end{align}
As expected, the velocity-dependent phase is proportional to the total area under the phase sensitivity function.

If we instead calculate the total interferometer phase using the inertial response function $h(t)$, we find:
\begin{align} \label{eq:response-limits}
    \Phi/k &= -\big[h(t)v(t) \big]_{-\infty}^{+\infty} + a\int_{-\infty}^{+\infty}h(t)\,\mathrm{d}t, \\
    &= \int_{-\infty}^{t_i}h(t_i)a\,\mathrm{d}t+ a\int_{t_i}^{+\infty}h(t)\,\mathrm{d}t, \\ \label{eq:open-response-function-scale-factor}
    &= h(t_i)v(t_i) + a\int_{t_i}^{+\infty}h(t)\,\mathrm{d}t.
\end{align}
We have assumed that at $t=-\infty$ the atoms are stationary (which is a reasonable assumption if atoms are released at rest from a trap prior to the interferometer). This allows us to set the first term on the right-hand side of Equation~\eqref{eq:response-limits} to zero and relate the integral of the response function before the interferometer starts at time $t_i$ to the initial velocity $v(t_i)$. We see from Equation~\eqref{eq:open-response-function-scale-factor} that - unlike in the phase sensitivity function calculation - the area under the response function only gives the scale factor if we take $v_0$ as the velocity of atoms at the start time of the pulse sequence.

Suppose instead we choose to define $t_0$ within the mirror pulse, e.g. at the origin $\tau_2^o$. In this case, we can rearrange Equation~\eqref{eq:open-response-function-scale-factor} using $v(t_i)=v_0+a(t_i-t_0)$ as follows:
\begin{align}
    \Phi/k&=v_0h(t_i)+a\bigg[\int_{t_i}^{+\infty}h(t)\,\mathrm{d}t+(t_i-t_0)h(t_i)  \bigg],\\
    &=v_0h(t_i)+a\int_{-\infty}^{+\infty}\tilde{h}(t)\,\mathrm{d}t.
\end{align}
In other words, the scale factor is in this case given by the area under a modified response function which is shifted vertically from $t_i$ to $t_0$ such that $\tilde{h}$ starts and ends at 0, as shown in Figure~\ref{fig:pulse-origin-diagram-open}. In this case, the scale factor is therefore approximately given by
\begin{align} \nonumber
    \mathcal{S} &\approx\frac{k}{2}\big[[T+(\tau_1-\tau^o_1)+\tau^o_2]^2 + [T+(\tau_2-\tau_2^o)+\tau^o_3]^2\big], \\ 
    &\approx k[T^2+T\tau_2+(m_3-m_1)T].
\end{align}
This expression is identical to Equation~\eqref{eq:approx-scale-factor} defined in the main text. We note that, in the case where $v_0$ is defined as the mean velocity during the mirror pulse, $\mathcal{S}$ does not depend to first-order upon $\tau_2^o$. The mirror pulse temporal origin still affects the velocity-dependent bias phase, however.

\bibliography{pulse_origin}

\providecommand{\noopsort}[1]{}\providecommand{\singleletter}[1]{#1}%
\begin{thebibliography}{56}%
\makeatletter
\providecommand \@ifxundefined [1]{%
 \@ifx{#1\undefined}
}%
\providecommand \@ifnum [1]{%
 \ifnum #1\expandafter \@firstoftwo
 \else \expandafter \@secondoftwo
 \fi
}%
\providecommand \@ifx [1]{%
 \ifx #1\expandafter \@firstoftwo
 \else \expandafter \@secondoftwo
 \fi
}%
\providecommand \natexlab [1]{#1}%
\providecommand \enquote  [1]{``#1''}%
\providecommand \bibnamefont  [1]{#1}%
\providecommand \bibfnamefont [1]{#1}%
\providecommand \citenamefont [1]{#1}%
\providecommand \href@noop [0]{\@secondoftwo}%
\providecommand \href [0]{\begingroup \@sanitize@url \@href}%
\providecommand \@href[1]{\@@startlink{#1}\@@href}%
\providecommand \@@href[1]{\endgroup#1\@@endlink}%
\providecommand \@sanitize@url [0]{\catcode `\\12\catcode `\$12\catcode `\&12\catcode `\#12\catcode `\^12\catcode `\_12\catcode `\%12\relax}%
\providecommand \@@startlink[1]{}%
\providecommand \@@endlink[0]{}%
\providecommand \url  [0]{\begingroup\@sanitize@url \@url }%
\providecommand \@url [1]{\endgroup\@href {#1}{\urlprefix }}%
\providecommand \urlprefix  [0]{URL }%
\providecommand \Eprint [0]{\href }%
\providecommand \doibase [0]{https://doi.org/}%
\providecommand \selectlanguage [0]{\@gobble}%
\providecommand \bibinfo  [0]{\@secondoftwo}%
\providecommand \bibfield  [0]{\@secondoftwo}%
\providecommand \translation [1]{[#1]}%
\providecommand \BibitemOpen [0]{}%
\providecommand \bibitemStop [0]{}%
\providecommand \bibitemNoStop [0]{.\EOS\space}%
\providecommand \EOS [0]{\spacefactor3000\relax}%
\providecommand \BibitemShut  [1]{\csname bibitem#1\endcsname}%
\let\auto@bib@innerbib\@empty
%</preamble>
\bibitem [{\citenamefont {Kasevich}\ and\ \citenamefont {Chu}(1991)}]{kasevich1991atomic}%
  \BibitemOpen
  \bibfield  {author} {\bibinfo {author} {\bibfnamefont {M.}~\bibnamefont {Kasevich}}\ and\ \bibinfo {author} {\bibfnamefont {S.}~\bibnamefont {Chu}},\ }\bibfield  {title} {\bibinfo {title} {{Atomic interferometry using stimulated Raman transitions}},\ }\href {https://doi.org/10.1103/PhysRevLett.67.181} {\bibfield  {journal} {\bibinfo  {journal} {Phys. Rev. Lett.}\ }\textbf {\bibinfo {volume} {67}},\ \bibinfo {pages} {181} (\bibinfo {year} {1991})}\BibitemShut {NoStop}%
\bibitem [{\citenamefont {Balasubramanian}\ \emph {et~al.}(2008)\citenamefont {Balasubramanian}, \citenamefont {Chan}, \citenamefont {Kolesov}, \citenamefont {Al-Hmoud}, \citenamefont {Tisler}, \citenamefont {Shin}, \citenamefont {Kim}, \citenamefont {Wojcik}, \citenamefont {Hemmer}, \citenamefont {Krueger} \emph {et~al.}}]{balasubramanian2008nanoscale}%
  \BibitemOpen
  \bibfield  {author} {\bibinfo {author} {\bibfnamefont {G.}~\bibnamefont {Balasubramanian}}, \bibinfo {author} {\bibfnamefont {I.}~\bibnamefont {Chan}}, \bibinfo {author} {\bibfnamefont {R.}~\bibnamefont {Kolesov}}, \bibinfo {author} {\bibfnamefont {M.}~\bibnamefont {Al-Hmoud}}, \bibinfo {author} {\bibfnamefont {J.}~\bibnamefont {Tisler}}, \bibinfo {author} {\bibfnamefont {C.}~\bibnamefont {Shin}}, \bibinfo {author} {\bibfnamefont {C.}~\bibnamefont {Kim}}, \bibinfo {author} {\bibfnamefont {A.}~\bibnamefont {Wojcik}}, \bibinfo {author} {\bibfnamefont {P.~R.}\ \bibnamefont {Hemmer}}, \bibinfo {author} {\bibfnamefont {A.}~\bibnamefont {Krueger}}, \emph {et~al.},\ }\bibfield  {title} {\bibinfo {title} {{Nanoscale imaging magnetometry with diamond spins under ambient conditions}},\ }\href {https://doi.org/10.1038/nature07278} {\bibfield  {journal} {\bibinfo  {journal} {Nature}\ }\textbf {\bibinfo {volume} {455}},\ \bibinfo {pages} {648} (\bibinfo {year} {2008})}\BibitemShut {NoStop}%
\bibitem [{\citenamefont {DiVincenzo}(2000)}]{divincenzo2000physical}%
  \BibitemOpen
  \bibfield  {author} {\bibinfo {author} {\bibfnamefont {D.~P.}\ \bibnamefont {DiVincenzo}},\ }\bibfield  {title} {\bibinfo {title} {{The physical implementation of quantum computation}},\ }\href {https://doi.org/10.1002/1521-3978(200009)48:9/11%3C771::AID-PROP771%3E3.0.CO;2-E} {\bibfield  {journal} {\bibinfo  {journal} {Fortschr. Phys.}\ }\textbf {\bibinfo {volume} {48}},\ \bibinfo {pages} {771} (\bibinfo {year} {2000})}\BibitemShut {NoStop}%
\bibitem [{\citenamefont {Steane}(1996)}]{steane1996error}%
  \BibitemOpen
  \bibfield  {author} {\bibinfo {author} {\bibfnamefont {A.~M.}\ \bibnamefont {Steane}},\ }\bibfield  {title} {\bibinfo {title} {{Error correcting codes in quantum theory}},\ }\href {https://doi.org/10.1103/PhysRevLett.77.793} {\bibfield  {journal} {\bibinfo  {journal} {Phys. Rev. Lett.}\ }\textbf {\bibinfo {volume} {77}},\ \bibinfo {pages} {793} (\bibinfo {year} {1996})}\BibitemShut {NoStop}%
\bibitem [{\citenamefont {Glaser}\ \emph {et~al.}(2015)\citenamefont {Glaser}, \citenamefont {Boscain}, \citenamefont {Calarco}, \citenamefont {Koch}, \citenamefont {K{\"o}ckenberger}, \citenamefont {Kosloff}, \citenamefont {Kuprov}, \citenamefont {Luy}, \citenamefont {Schirmer}, \citenamefont {Schulte-Herbr{\"u}ggen} \emph {et~al.}}]{glaser2015training}%
  \BibitemOpen
  \bibfield  {author} {\bibinfo {author} {\bibfnamefont {S.~J.}\ \bibnamefont {Glaser}}, \bibinfo {author} {\bibfnamefont {U.}~\bibnamefont {Boscain}}, \bibinfo {author} {\bibfnamefont {T.}~\bibnamefont {Calarco}}, \bibinfo {author} {\bibfnamefont {C.~P.}\ \bibnamefont {Koch}}, \bibinfo {author} {\bibfnamefont {W.}~\bibnamefont {K{\"o}ckenberger}}, \bibinfo {author} {\bibfnamefont {R.}~\bibnamefont {Kosloff}}, \bibinfo {author} {\bibfnamefont {I.}~\bibnamefont {Kuprov}}, \bibinfo {author} {\bibfnamefont {B.}~\bibnamefont {Luy}}, \bibinfo {author} {\bibfnamefont {S.}~\bibnamefont {Schirmer}}, \bibinfo {author} {\bibfnamefont {T.}~\bibnamefont {Schulte-Herbr{\"u}ggen}}, \emph {et~al.},\ }\bibfield  {title} {\bibinfo {title} {{Training Schr{\"o}dinger’s cat: Quantum optimal control: Strategic report on current status, visions and goals for research in Europe}},\ }\href {https://doi.org/10.1140/epjd/e2015-60464-1} {\bibfield  {journal} {\bibinfo  {journal} {Eur. Phys. J. D}\ }\textbf {\bibinfo {volume} {69}},\
  \bibinfo {pages} {279} (\bibinfo {year} {2015})}\BibitemShut {NoStop}%
\bibitem [{\citenamefont {Barrett}\ \emph {et~al.}(2014)\citenamefont {Barrett}, \citenamefont {Geiger}, \citenamefont {Dutta}, \citenamefont {Meunier}, \citenamefont {Canuel}, \citenamefont {Gauguet}, \citenamefont {Bouyer},\ and\ \citenamefont {Landragin}}]{barrett2014sagnac}%
  \BibitemOpen
  \bibfield  {author} {\bibinfo {author} {\bibfnamefont {B.}~\bibnamefont {Barrett}}, \bibinfo {author} {\bibfnamefont {R.}~\bibnamefont {Geiger}}, \bibinfo {author} {\bibfnamefont {I.}~\bibnamefont {Dutta}}, \bibinfo {author} {\bibfnamefont {M.}~\bibnamefont {Meunier}}, \bibinfo {author} {\bibfnamefont {B.}~\bibnamefont {Canuel}}, \bibinfo {author} {\bibfnamefont {A.}~\bibnamefont {Gauguet}}, \bibinfo {author} {\bibfnamefont {P.}~\bibnamefont {Bouyer}},\ and\ \bibinfo {author} {\bibfnamefont {A.}~\bibnamefont {Landragin}},\ }\bibfield  {title} {\bibinfo {title} {{The Sagnac effect: 20 years of development in matter-wave interferometry}},\ }\href {https://doi.org/10.1016/j.crhy.2014.10.009} {\bibfield  {journal} {\bibinfo  {journal} {C. R. Phys.}\ }\textbf {\bibinfo {volume} {15}},\ \bibinfo {pages} {875} (\bibinfo {year} {2014})}\BibitemShut {NoStop}%
\bibitem [{\citenamefont {Canuel}\ \emph {et~al.}(2006)\citenamefont {Canuel}, \citenamefont {Leduc}, \citenamefont {Holleville}, \citenamefont {Gauguet}, \citenamefont {Fils}, \citenamefont {Virdis}, \citenamefont {Clairon}, \citenamefont {Dimarcq}, \citenamefont {Bord{\'e}}, \citenamefont {Landragin} \emph {et~al.}}]{canuel2006six}%
  \BibitemOpen
  \bibfield  {author} {\bibinfo {author} {\bibfnamefont {B.}~\bibnamefont {Canuel}}, \bibinfo {author} {\bibfnamefont {F.}~\bibnamefont {Leduc}}, \bibinfo {author} {\bibfnamefont {D.}~\bibnamefont {Holleville}}, \bibinfo {author} {\bibfnamefont {A.}~\bibnamefont {Gauguet}}, \bibinfo {author} {\bibfnamefont {J.}~\bibnamefont {Fils}}, \bibinfo {author} {\bibfnamefont {A.}~\bibnamefont {Virdis}}, \bibinfo {author} {\bibfnamefont {A.}~\bibnamefont {Clairon}}, \bibinfo {author} {\bibfnamefont {N.}~\bibnamefont {Dimarcq}}, \bibinfo {author} {\bibfnamefont {C.~J.}\ \bibnamefont {Bord{\'e}}}, \bibinfo {author} {\bibfnamefont {A.}~\bibnamefont {Landragin}}, \emph {et~al.},\ }\bibfield  {title} {\bibinfo {title} {{Six-axis inertial sensor using cold-atom interferometry}},\ }\href {https://doi.org/10.1103/PhysRevLett.97.010402} {\bibfield  {journal} {\bibinfo  {journal} {Phys. Rev. Lett.}\ }\textbf {\bibinfo {volume} {97}},\ \bibinfo {pages} {010402} (\bibinfo {year} {2006})}\BibitemShut {NoStop}%
\bibitem [{\citenamefont {d’Armagnac~de Castanet}\ \emph {et~al.}(2024)\citenamefont {d’Armagnac~de Castanet}, \citenamefont {Des~Cognets}, \citenamefont {Arguel}, \citenamefont {Templier}, \citenamefont {Jarlaud}, \citenamefont {M{\'e}noret}, \citenamefont {Desruelle}, \citenamefont {Bouyer},\ and\ \citenamefont {Battelier}}]{d2024atom}%
  \BibitemOpen
  \bibfield  {author} {\bibinfo {author} {\bibfnamefont {Q.}~\bibnamefont {d’Armagnac~de Castanet}}, \bibinfo {author} {\bibfnamefont {C.}~\bibnamefont {Des~Cognets}}, \bibinfo {author} {\bibfnamefont {R.}~\bibnamefont {Arguel}}, \bibinfo {author} {\bibfnamefont {S.}~\bibnamefont {Templier}}, \bibinfo {author} {\bibfnamefont {V.}~\bibnamefont {Jarlaud}}, \bibinfo {author} {\bibfnamefont {V.}~\bibnamefont {M{\'e}noret}}, \bibinfo {author} {\bibfnamefont {B.}~\bibnamefont {Desruelle}}, \bibinfo {author} {\bibfnamefont {P.}~\bibnamefont {Bouyer}},\ and\ \bibinfo {author} {\bibfnamefont {B.}~\bibnamefont {Battelier}},\ }\bibfield  {title} {\bibinfo {title} {{Atom interferometry at arbitrary orientations and rotation rates}},\ }\href {https://doi.org/10.1038/s41467-024-50804-0} {\bibfield  {journal} {\bibinfo  {journal} {Nat. Commun.}\ }\textbf {\bibinfo {volume} {15}},\ \bibinfo {pages} {6406} (\bibinfo {year} {2024})}\BibitemShut {NoStop}%
\bibitem [{\citenamefont {Templier}\ \emph {et~al.}(2022)\citenamefont {Templier}, \citenamefont {Cheiney}, \citenamefont {d’Armagnac~de Castanet}, \citenamefont {Gouraud}, \citenamefont {Porte}, \citenamefont {Napolitano}, \citenamefont {Bouyer}, \citenamefont {Battelier},\ and\ \citenamefont {Barrett}}]{templier2022tracking}%
  \BibitemOpen
  \bibfield  {author} {\bibinfo {author} {\bibfnamefont {S.}~\bibnamefont {Templier}}, \bibinfo {author} {\bibfnamefont {P.}~\bibnamefont {Cheiney}}, \bibinfo {author} {\bibfnamefont {Q.}~\bibnamefont {d’Armagnac~de Castanet}}, \bibinfo {author} {\bibfnamefont {B.}~\bibnamefont {Gouraud}}, \bibinfo {author} {\bibfnamefont {H.}~\bibnamefont {Porte}}, \bibinfo {author} {\bibfnamefont {F.}~\bibnamefont {Napolitano}}, \bibinfo {author} {\bibfnamefont {P.}~\bibnamefont {Bouyer}}, \bibinfo {author} {\bibfnamefont {B.}~\bibnamefont {Battelier}},\ and\ \bibinfo {author} {\bibfnamefont {B.}~\bibnamefont {Barrett}},\ }\bibfield  {title} {\bibinfo {title} {{Tracking the vector acceleration with a hybrid quantum accelerometer triad}},\ }\href {https://doi.org/https://doi.org/10.1126/sciadv.add3854} {\bibfield  {journal} {\bibinfo  {journal} {Sci. Adv.}\ }\textbf {\bibinfo {volume} {8}},\ \bibinfo {pages} {eadd3854} (\bibinfo {year} {2022})}\BibitemShut {NoStop}%
\bibitem [{\citenamefont {Geiger}\ \emph {et~al.}(2011)\citenamefont {Geiger}, \citenamefont {M{\'e}noret}, \citenamefont {Stern}, \citenamefont {Zahzam}, \citenamefont {Cheinet}, \citenamefont {Battelier}, \citenamefont {Villing}, \citenamefont {Moron}, \citenamefont {Lours}, \citenamefont {Bidel} \emph {et~al.}}]{geiger2011detecting}%
  \BibitemOpen
  \bibfield  {author} {\bibinfo {author} {\bibfnamefont {R.}~\bibnamefont {Geiger}}, \bibinfo {author} {\bibfnamefont {V.}~\bibnamefont {M{\'e}noret}}, \bibinfo {author} {\bibfnamefont {G.}~\bibnamefont {Stern}}, \bibinfo {author} {\bibfnamefont {N.}~\bibnamefont {Zahzam}}, \bibinfo {author} {\bibfnamefont {P.}~\bibnamefont {Cheinet}}, \bibinfo {author} {\bibfnamefont {B.}~\bibnamefont {Battelier}}, \bibinfo {author} {\bibfnamefont {A.}~\bibnamefont {Villing}}, \bibinfo {author} {\bibfnamefont {F.}~\bibnamefont {Moron}}, \bibinfo {author} {\bibfnamefont {M.}~\bibnamefont {Lours}}, \bibinfo {author} {\bibfnamefont {Y.}~\bibnamefont {Bidel}}, \emph {et~al.},\ }\bibfield  {title} {\bibinfo {title} {{Detecting inertial effects with airborne matter-wave interferometry}},\ }\href {https://doi.org/10.1038/ncomms1479} {\bibfield  {journal} {\bibinfo  {journal} {Nat. Commun.}\ }\textbf {\bibinfo {volume} {2}},\ \bibinfo {pages} {474} (\bibinfo {year} {2011})}\BibitemShut {NoStop}%
\bibitem [{\citenamefont {Mcguirk}\ \emph {et~al.}(2002)\citenamefont {Mcguirk}, \citenamefont {Foster}, \citenamefont {Fixler}, \citenamefont {Snadden},\ and\ \citenamefont {Kasevich}}]{mcguirk2002sensitive}%
  \BibitemOpen
  \bibfield  {author} {\bibinfo {author} {\bibfnamefont {J.~M.}\ \bibnamefont {Mcguirk}}, \bibinfo {author} {\bibfnamefont {G.}~\bibnamefont {Foster}}, \bibinfo {author} {\bibfnamefont {J.}~\bibnamefont {Fixler}}, \bibinfo {author} {\bibfnamefont {M.}~\bibnamefont {Snadden}},\ and\ \bibinfo {author} {\bibfnamefont {M.}~\bibnamefont {Kasevich}},\ }\bibfield  {title} {\bibinfo {title} {{Sensitive absolute-gravity gradiometry using atom interferometry}},\ }\href {https://doi.org/10.1103/PhysRevA.65.033608} {\bibfield  {journal} {\bibinfo  {journal} {Phys. Rev. A}\ }\textbf {\bibinfo {volume} {65}},\ \bibinfo {pages} {033608} (\bibinfo {year} {2002})}\BibitemShut {NoStop}%
\bibitem [{\citenamefont {Bidel}\ \emph {et~al.}(2018)\citenamefont {Bidel}, \citenamefont {Zahzam}, \citenamefont {Blanchard}, \citenamefont {Bonnin}, \citenamefont {Cadoret}, \citenamefont {Bresson}, \citenamefont {Rouxel},\ and\ \citenamefont {Lequentrec-Lalancette}}]{bidel2018absolute}%
  \BibitemOpen
  \bibfield  {author} {\bibinfo {author} {\bibfnamefont {Y.}~\bibnamefont {Bidel}}, \bibinfo {author} {\bibfnamefont {N.}~\bibnamefont {Zahzam}}, \bibinfo {author} {\bibfnamefont {C.}~\bibnamefont {Blanchard}}, \bibinfo {author} {\bibfnamefont {A.}~\bibnamefont {Bonnin}}, \bibinfo {author} {\bibfnamefont {M.}~\bibnamefont {Cadoret}}, \bibinfo {author} {\bibfnamefont {A.}~\bibnamefont {Bresson}}, \bibinfo {author} {\bibfnamefont {D.}~\bibnamefont {Rouxel}},\ and\ \bibinfo {author} {\bibfnamefont {M.}~\bibnamefont {Lequentrec-Lalancette}},\ }\bibfield  {title} {\bibinfo {title} {{Absolute marine gravimetry with matter-wave interferometry}},\ }\href {https://doi.org/10.1038/s41467-018-03040-2} {\bibfield  {journal} {\bibinfo  {journal} {Nat. Commun.}\ }\textbf {\bibinfo {volume} {9}},\ \bibinfo {pages} {627} (\bibinfo {year} {2018})}\BibitemShut {NoStop}%
\bibitem [{\citenamefont {Stray}\ \emph {et~al.}(2022)\citenamefont {Stray}, \citenamefont {Lamb}, \citenamefont {Kaushik}, \citenamefont {Vovrosh}, \citenamefont {Rodgers}, \citenamefont {Winch}, \citenamefont {Hayati}, \citenamefont {Boddice}, \citenamefont {Stabrawa}, \citenamefont {Niggebaum} \emph {et~al.}}]{stray2022quantum}%
  \BibitemOpen
  \bibfield  {author} {\bibinfo {author} {\bibfnamefont {B.}~\bibnamefont {Stray}}, \bibinfo {author} {\bibfnamefont {A.}~\bibnamefont {Lamb}}, \bibinfo {author} {\bibfnamefont {A.}~\bibnamefont {Kaushik}}, \bibinfo {author} {\bibfnamefont {J.}~\bibnamefont {Vovrosh}}, \bibinfo {author} {\bibfnamefont {A.}~\bibnamefont {Rodgers}}, \bibinfo {author} {\bibfnamefont {J.}~\bibnamefont {Winch}}, \bibinfo {author} {\bibfnamefont {F.}~\bibnamefont {Hayati}}, \bibinfo {author} {\bibfnamefont {D.}~\bibnamefont {Boddice}}, \bibinfo {author} {\bibfnamefont {A.}~\bibnamefont {Stabrawa}}, \bibinfo {author} {\bibfnamefont {A.}~\bibnamefont {Niggebaum}}, \emph {et~al.},\ }\bibfield  {title} {\bibinfo {title} {{Quantum sensing for gravity cartography}},\ }\href {https://doi.org/10.1038/s41586-021-04315-3} {\bibfield  {journal} {\bibinfo  {journal} {Nature}\ }\textbf {\bibinfo {volume} {602}},\ \bibinfo {pages} {590} (\bibinfo {year} {2022})}\BibitemShut {NoStop}%
\bibitem [{\citenamefont {Narducci}\ \emph {et~al.}(2022)\citenamefont {Narducci}, \citenamefont {Black},\ and\ \citenamefont {Burke}}]{narducci2022advances}%
  \BibitemOpen
  \bibfield  {author} {\bibinfo {author} {\bibfnamefont {F.~A.}\ \bibnamefont {Narducci}}, \bibinfo {author} {\bibfnamefont {A.~T.}\ \bibnamefont {Black}},\ and\ \bibinfo {author} {\bibfnamefont {J.~H.}\ \bibnamefont {Burke}},\ }\bibfield  {title} {\bibinfo {title} {{Advances toward fieldable atom interferometers}},\ }\href {https://doi.org/10.1080/23746149.2021.1946426} {\bibfield  {journal} {\bibinfo  {journal} {Adv. Phys. X}\ }\textbf {\bibinfo {volume} {7}},\ \bibinfo {pages} {1946426} (\bibinfo {year} {2022})}\BibitemShut {NoStop}%
\bibitem [{\citenamefont {Phillips}\ \emph {et~al.}(2022)\citenamefont {Phillips}, \citenamefont {Wright}, \citenamefont {Riou}, \citenamefont {Maddox}, \citenamefont {Maskell},\ and\ \citenamefont {Ralph}}]{phillips2022position}%
  \BibitemOpen
  \bibfield  {author} {\bibinfo {author} {\bibfnamefont {A.~M.}\ \bibnamefont {Phillips}}, \bibinfo {author} {\bibfnamefont {M.~J.}\ \bibnamefont {Wright}}, \bibinfo {author} {\bibfnamefont {I.}~\bibnamefont {Riou}}, \bibinfo {author} {\bibfnamefont {S.}~\bibnamefont {Maddox}}, \bibinfo {author} {\bibfnamefont {S.}~\bibnamefont {Maskell}},\ and\ \bibinfo {author} {\bibfnamefont {J.~F.}\ \bibnamefont {Ralph}},\ }\bibfield  {title} {\bibinfo {title} {{Position fixing with cold atom gravity gradiometers}},\ }\href {https://doi.org/10.1116/5.0095677} {\bibfield  {journal} {\bibinfo  {journal} {AVS Quantum Sci.}\ }\textbf {\bibinfo {volume} {4}} (\bibinfo {year} {2022})}\BibitemShut {NoStop}%
\bibitem [{\citenamefont {Badurina}\ \emph {et~al.}(2020)\citenamefont {Badurina}, \citenamefont {Bentine}, \citenamefont {Blas}, \citenamefont {Bongs}, \citenamefont {Bortoletto}, \citenamefont {Bowcock}, \citenamefont {Bridges}, \citenamefont {Bowden}, \citenamefont {Buchmueller}, \citenamefont {Burrage} \emph {et~al.}}]{badurina2020aion}%
  \BibitemOpen
  \bibfield  {author} {\bibinfo {author} {\bibfnamefont {L.}~\bibnamefont {Badurina}}, \bibinfo {author} {\bibfnamefont {E.}~\bibnamefont {Bentine}}, \bibinfo {author} {\bibfnamefont {D.}~\bibnamefont {Blas}}, \bibinfo {author} {\bibfnamefont {K.}~\bibnamefont {Bongs}}, \bibinfo {author} {\bibfnamefont {D.}~\bibnamefont {Bortoletto}}, \bibinfo {author} {\bibfnamefont {T.}~\bibnamefont {Bowcock}}, \bibinfo {author} {\bibfnamefont {K.}~\bibnamefont {Bridges}}, \bibinfo {author} {\bibfnamefont {W.}~\bibnamefont {Bowden}}, \bibinfo {author} {\bibfnamefont {O.}~\bibnamefont {Buchmueller}}, \bibinfo {author} {\bibfnamefont {C.}~\bibnamefont {Burrage}}, \emph {et~al.},\ }\bibfield  {title} {\bibinfo {title} {{AION: an atom interferometer observatory and network}},\ }\href {https://doi.org/10.1088/1475-7516/2020/05/011} {\bibfield  {journal} {\bibinfo  {journal} {J. Cosmol. Astropart. Phys.}\ }\textbf {\bibinfo {volume} {2020}}\bibinfo  {number} { (05)},\ \bibinfo {pages} {011}}\BibitemShut {NoStop}%
\bibitem [{\citenamefont {Canuel}\ \emph {et~al.}(2018)\citenamefont {Canuel}, \citenamefont {Bertoldi}, \citenamefont {Amand}, \citenamefont {Pozzo~di Borgo}, \citenamefont {Chantrait}, \citenamefont {Danquigny}, \citenamefont {Dovale~{\'A}lvarez}, \citenamefont {Fang}, \citenamefont {Freise}, \citenamefont {Geiger} \emph {et~al.}}]{canuel2018exploring}%
  \BibitemOpen
\bibfield  {number} {  }\bibfield  {author} {\bibinfo {author} {\bibfnamefont {B.}~\bibnamefont {Canuel}}, \bibinfo {author} {\bibfnamefont {A.}~\bibnamefont {Bertoldi}}, \bibinfo {author} {\bibfnamefont {L.}~\bibnamefont {Amand}}, \bibinfo {author} {\bibfnamefont {E.}~\bibnamefont {Pozzo~di Borgo}}, \bibinfo {author} {\bibfnamefont {T.}~\bibnamefont {Chantrait}}, \bibinfo {author} {\bibfnamefont {C.}~\bibnamefont {Danquigny}}, \bibinfo {author} {\bibfnamefont {M.}~\bibnamefont {Dovale~{\'A}lvarez}}, \bibinfo {author} {\bibfnamefont {B.}~\bibnamefont {Fang}}, \bibinfo {author} {\bibfnamefont {A.}~\bibnamefont {Freise}}, \bibinfo {author} {\bibfnamefont {R.}~\bibnamefont {Geiger}}, \emph {et~al.},\ }\bibfield  {title} {\bibinfo {title} {{Exploring gravity with the MIGA large scale atom interferometer}},\ }\href {https://doi.org/https://doi.org/10.1038/s41598-018-32165-z} {\bibfield  {journal} {\bibinfo  {journal} {Sci. Rep.}\ }\textbf {\bibinfo {volume} {8}},\ \bibinfo {pages} {14064} (\bibinfo {year}
  {2018})}\BibitemShut {NoStop}%
\bibitem [{\citenamefont {Hamilton}\ \emph {et~al.}(2015)\citenamefont {Hamilton}, \citenamefont {Jaffe}, \citenamefont {Haslinger}, \citenamefont {Simmons}, \citenamefont {M{\"u}ller},\ and\ \citenamefont {Khoury}}]{hamilton2015atom}%
  \BibitemOpen
  \bibfield  {author} {\bibinfo {author} {\bibfnamefont {P.}~\bibnamefont {Hamilton}}, \bibinfo {author} {\bibfnamefont {M.}~\bibnamefont {Jaffe}}, \bibinfo {author} {\bibfnamefont {P.}~\bibnamefont {Haslinger}}, \bibinfo {author} {\bibfnamefont {Q.}~\bibnamefont {Simmons}}, \bibinfo {author} {\bibfnamefont {H.}~\bibnamefont {M{\"u}ller}},\ and\ \bibinfo {author} {\bibfnamefont {J.}~\bibnamefont {Khoury}},\ }\bibfield  {title} {\bibinfo {title} {{Atom-interferometry constraints on dark energy}},\ }\href {https://doi.org/10.1126/science.aaa8883} {\bibfield  {journal} {\bibinfo  {journal} {Science}\ }\textbf {\bibinfo {volume} {349}},\ \bibinfo {pages} {849} (\bibinfo {year} {2015})}\BibitemShut {NoStop}%
\bibitem [{\citenamefont {Fang}\ \emph {et~al.}(2018)\citenamefont {Fang}, \citenamefont {Mielec}, \citenamefont {Savoie}, \citenamefont {Altorio}, \citenamefont {Landragin},\ and\ \citenamefont {Geiger}}]{fang2018improving}%
  \BibitemOpen
  \bibfield  {author} {\bibinfo {author} {\bibfnamefont {B.}~\bibnamefont {Fang}}, \bibinfo {author} {\bibfnamefont {N.}~\bibnamefont {Mielec}}, \bibinfo {author} {\bibfnamefont {D.}~\bibnamefont {Savoie}}, \bibinfo {author} {\bibfnamefont {M.}~\bibnamefont {Altorio}}, \bibinfo {author} {\bibfnamefont {A.}~\bibnamefont {Landragin}},\ and\ \bibinfo {author} {\bibfnamefont {R.}~\bibnamefont {Geiger}},\ }\bibfield  {title} {\bibinfo {title} {Improving the phase response of an atom interferometer by means of temporal pulse shaping},\ }\href {https://doi.org/https://doi.org/10.1088/1367-2630/aaa37c} {\bibfield  {journal} {\bibinfo  {journal} {New J. Phys.}\ }\textbf {\bibinfo {volume} {20}},\ \bibinfo {pages} {023020} (\bibinfo {year} {2018})}\BibitemShut {NoStop}%
\bibitem [{\citenamefont {Bonnin}\ \emph {et~al.}(2015)\citenamefont {Bonnin}, \citenamefont {Zahzam}, \citenamefont {Bidel},\ and\ \citenamefont {Bresson}}]{bonnin2015characterization}%
  \BibitemOpen
  \bibfield  {author} {\bibinfo {author} {\bibfnamefont {A.}~\bibnamefont {Bonnin}}, \bibinfo {author} {\bibfnamefont {N.}~\bibnamefont {Zahzam}}, \bibinfo {author} {\bibfnamefont {Y.}~\bibnamefont {Bidel}},\ and\ \bibinfo {author} {\bibfnamefont {A.}~\bibnamefont {Bresson}},\ }\bibfield  {title} {\bibinfo {title} {Characterization of a simultaneous dual-species atom interferometer for a quantum test of the weak equivalence principle},\ }\href {https://doi.org/https://doi.org/10.1103/PhysRevA.92.023626} {\bibfield  {journal} {\bibinfo  {journal} {Phys. Rev. A}\ }\textbf {\bibinfo {volume} {92}},\ \bibinfo {pages} {023626} (\bibinfo {year} {2015})}\BibitemShut {NoStop}%
\bibitem [{\citenamefont {Gillot}\ \emph {et~al.}(2016)\citenamefont {Gillot}, \citenamefont {Cheng}, \citenamefont {Merlet},\ and\ \citenamefont {Pereira Dos~Santos}}]{gillot2016limits}%
  \BibitemOpen
  \bibfield  {author} {\bibinfo {author} {\bibfnamefont {P.}~\bibnamefont {Gillot}}, \bibinfo {author} {\bibfnamefont {B.}~\bibnamefont {Cheng}}, \bibinfo {author} {\bibfnamefont {S.}~\bibnamefont {Merlet}},\ and\ \bibinfo {author} {\bibfnamefont {F.}~\bibnamefont {Pereira Dos~Santos}},\ }\bibfield  {title} {\bibinfo {title} {{Limits to the symmetry of a Mach-Zehnder-type atom interferometer}},\ }\href {https://doi.org/https://doi.org/10.1103/PhysRevA.93.013609} {\bibfield  {journal} {\bibinfo  {journal} {Phys. Rev. A}\ }\textbf {\bibinfo {volume} {93}},\ \bibinfo {pages} {013609} (\bibinfo {year} {2016})}\BibitemShut {NoStop}%
\bibitem [{\citenamefont {Barrett}\ \emph {et~al.}(2016)\citenamefont {Barrett}, \citenamefont {Antoni-Micollier}, \citenamefont {Chichet}, \citenamefont {Battelier}, \citenamefont {L{\'e}veque}, \citenamefont {Landragin},\ and\ \citenamefont {Bouyer}}]{barrett2016dual}%
  \BibitemOpen
  \bibfield  {author} {\bibinfo {author} {\bibfnamefont {B.}~\bibnamefont {Barrett}}, \bibinfo {author} {\bibfnamefont {L.}~\bibnamefont {Antoni-Micollier}}, \bibinfo {author} {\bibfnamefont {L.}~\bibnamefont {Chichet}}, \bibinfo {author} {\bibfnamefont {B.}~\bibnamefont {Battelier}}, \bibinfo {author} {\bibfnamefont {T.}~\bibnamefont {L{\'e}veque}}, \bibinfo {author} {\bibfnamefont {A.}~\bibnamefont {Landragin}},\ and\ \bibinfo {author} {\bibfnamefont {P.}~\bibnamefont {Bouyer}},\ }\bibfield  {title} {\bibinfo {title} {{Dual matter-wave inertial sensors in weightlessness}},\ }\href {https://doi.org/10.1038/ncomms13786} {\bibfield  {journal} {\bibinfo  {journal} {Nat. Commun.}\ }\textbf {\bibinfo {volume} {7}},\ \bibinfo {pages} {13786} (\bibinfo {year} {2016})}\BibitemShut {NoStop}%
\bibitem [{\citenamefont {Dedes}\ \emph {et~al.}(2023)\citenamefont {Dedes}, \citenamefont {Saywell}, \citenamefont {Carey}, \citenamefont {Kuprov},\ and\ \citenamefont {Freegarde}}]{dedes2023optimizing}%
  \BibitemOpen
  \bibfield  {author} {\bibinfo {author} {\bibfnamefont {N.}~\bibnamefont {Dedes}}, \bibinfo {author} {\bibfnamefont {J.}~\bibnamefont {Saywell}}, \bibinfo {author} {\bibfnamefont {M.}~\bibnamefont {Carey}}, \bibinfo {author} {\bibfnamefont {I.}~\bibnamefont {Kuprov}},\ and\ \bibinfo {author} {\bibfnamefont {T.}~\bibnamefont {Freegarde}},\ }\bibfield  {title} {\bibinfo {title} {{Optimizing beam-splitter pulses for atom interferometry: A geometric approach}},\ }\href {https://doi.org/https://doi.org/10.1103/PhysRevA.108.053319} {\bibfield  {journal} {\bibinfo  {journal} {Phys. Rev. A}\ }\textbf {\bibinfo {volume} {108}},\ \bibinfo {pages} {053319} (\bibinfo {year} {2023})}\BibitemShut {NoStop}%
\bibitem [{\citenamefont {Szigeti}\ \emph {et~al.}(2012)\citenamefont {Szigeti}, \citenamefont {Debs}, \citenamefont {Hope}, \citenamefont {Robins},\ and\ \citenamefont {Close}}]{szigeti2012momentum}%
  \BibitemOpen
  \bibfield  {author} {\bibinfo {author} {\bibfnamefont {S.~S.}\ \bibnamefont {Szigeti}}, \bibinfo {author} {\bibfnamefont {J.~E.}\ \bibnamefont {Debs}}, \bibinfo {author} {\bibfnamefont {J.~J.}\ \bibnamefont {Hope}}, \bibinfo {author} {\bibfnamefont {N.~P.}\ \bibnamefont {Robins}},\ and\ \bibinfo {author} {\bibfnamefont {J.~D.}\ \bibnamefont {Close}},\ }\bibfield  {title} {\bibinfo {title} {{Why momentum width matters for atom interferometry with Bragg pulses}},\ }\href {https://doi.org/10.1088/1367-2630/14/2/023009} {\bibfield  {journal} {\bibinfo  {journal} {New J. Phys.}\ }\textbf {\bibinfo {volume} {14}},\ \bibinfo {pages} {023009} (\bibinfo {year} {2012})}\BibitemShut {NoStop}%
\bibitem [{\citenamefont {Saywell}\ \emph {et~al.}(2018)\citenamefont {Saywell}, \citenamefont {Kuprov}, \citenamefont {Goodwin}, \citenamefont {Carey},\ and\ \citenamefont {Freegarde}}]{saywell2018optimal}%
  \BibitemOpen
  \bibfield  {author} {\bibinfo {author} {\bibfnamefont {J.~C.}\ \bibnamefont {Saywell}}, \bibinfo {author} {\bibfnamefont {I.}~\bibnamefont {Kuprov}}, \bibinfo {author} {\bibfnamefont {D.}~\bibnamefont {Goodwin}}, \bibinfo {author} {\bibfnamefont {M.}~\bibnamefont {Carey}},\ and\ \bibinfo {author} {\bibfnamefont {T.}~\bibnamefont {Freegarde}},\ }\bibfield  {title} {\bibinfo {title} {{Optimal control of mirror pulses for cold-atom interferometry}},\ }\href {https://doi.org/https://doi.org/10.1103/PhysRevA.98.023625} {\bibfield  {journal} {\bibinfo  {journal} {Phys. Rev. A}\ }\textbf {\bibinfo {volume} {98}},\ \bibinfo {pages} {023625} (\bibinfo {year} {2018})}\BibitemShut {NoStop}%
\bibitem [{\citenamefont {Saywell}\ \emph {et~al.}(2020{\natexlab{a}})\citenamefont {Saywell}, \citenamefont {Carey}, \citenamefont {Belal}, \citenamefont {Kuprov},\ and\ \citenamefont {Freegarde}}]{saywell2020optimal}%
  \BibitemOpen
  \bibfield  {author} {\bibinfo {author} {\bibfnamefont {J.}~\bibnamefont {Saywell}}, \bibinfo {author} {\bibfnamefont {M.}~\bibnamefont {Carey}}, \bibinfo {author} {\bibfnamefont {M.}~\bibnamefont {Belal}}, \bibinfo {author} {\bibfnamefont {I.}~\bibnamefont {Kuprov}},\ and\ \bibinfo {author} {\bibfnamefont {T.}~\bibnamefont {Freegarde}},\ }\bibfield  {title} {\bibinfo {title} {{Optimal control of Raman pulse sequences for atom interferometry}},\ }\href {https://doi.org/https://doi.org/10.1088/1361-6455/ab6df6} {\bibfield  {journal} {\bibinfo  {journal} {J. Phys. B: At. Mol. Opt. Phys.}\ }\textbf {\bibinfo {volume} {53}},\ \bibinfo {pages} {085006} (\bibinfo {year} {2020}{\natexlab{a}})}\BibitemShut {NoStop}%
\bibitem [{\citenamefont {Louie}\ \emph {et~al.}(2023)\citenamefont {Louie}, \citenamefont {Chen}, \citenamefont {Deshpande},\ and\ \citenamefont {Kovachy}}]{louie2023robust}%
  \BibitemOpen
  \bibfield  {author} {\bibinfo {author} {\bibfnamefont {G.}~\bibnamefont {Louie}}, \bibinfo {author} {\bibfnamefont {Z.}~\bibnamefont {Chen}}, \bibinfo {author} {\bibfnamefont {T.}~\bibnamefont {Deshpande}},\ and\ \bibinfo {author} {\bibfnamefont {T.}~\bibnamefont {Kovachy}},\ }\bibfield  {title} {\bibinfo {title} {{Robust atom optics for Bragg atom interferometry}},\ }\href {https://doi.org/https://doi.org/10.1088/1367-2630/aceb15} {\bibfield  {journal} {\bibinfo  {journal} {New J. Phys.}\ }\textbf {\bibinfo {volume} {25}},\ \bibinfo {pages} {083017} (\bibinfo {year} {2023})}\BibitemShut {NoStop}%
\bibitem [{\citenamefont {Chen}\ \emph {et~al.}(2023)\citenamefont {Chen}, \citenamefont {Louie}, \citenamefont {Wang}, \citenamefont {Deshpande},\ and\ \citenamefont {Kovachy}}]{chen2023enhancing}%
  \BibitemOpen
  \bibfield  {author} {\bibinfo {author} {\bibfnamefont {Z.}~\bibnamefont {Chen}}, \bibinfo {author} {\bibfnamefont {G.}~\bibnamefont {Louie}}, \bibinfo {author} {\bibfnamefont {Y.}~\bibnamefont {Wang}}, \bibinfo {author} {\bibfnamefont {T.}~\bibnamefont {Deshpande}},\ and\ \bibinfo {author} {\bibfnamefont {T.}~\bibnamefont {Kovachy}},\ }\bibfield  {title} {\bibinfo {title} {{Enhancing strontium clock atom interferometry using quantum optimal control}},\ }\href {https://doi.org/10.1103/PhysRevA.107.063302} {\bibfield  {journal} {\bibinfo  {journal} {Phys. Rev. A}\ }\textbf {\bibinfo {volume} {107}},\ \bibinfo {pages} {063302} (\bibinfo {year} {2023})}\BibitemShut {NoStop}%
\bibitem [{\citenamefont {Saywell}\ \emph {et~al.}(2020{\natexlab{b}})\citenamefont {Saywell}, \citenamefont {Carey}, \citenamefont {Kuprov},\ and\ \citenamefont {Freegarde}}]{saywell2020biselective}%
  \BibitemOpen
  \bibfield  {author} {\bibinfo {author} {\bibfnamefont {J.}~\bibnamefont {Saywell}}, \bibinfo {author} {\bibfnamefont {M.}~\bibnamefont {Carey}}, \bibinfo {author} {\bibfnamefont {I.}~\bibnamefont {Kuprov}},\ and\ \bibinfo {author} {\bibfnamefont {T.}~\bibnamefont {Freegarde}},\ }\bibfield  {title} {\bibinfo {title} {{Biselective pulses for large-area atom interferometry}},\ }\href {https://doi.org/10.1103/PhysRevA.101.063625} {\bibfield  {journal} {\bibinfo  {journal} {Phys. Rev. A}\ }\textbf {\bibinfo {volume} {101}},\ \bibinfo {pages} {063625} (\bibinfo {year} {2020}{\natexlab{b}})}\BibitemShut {NoStop}%
\bibitem [{\citenamefont {Lahuerta}\ \emph {et~al.}(2025)\citenamefont {Lahuerta}, \citenamefont {Kirsten-Siem{\ss}}, \citenamefont {Hammerer},\ and\ \citenamefont {Gaaloul}}]{lahuerta2025diffraction}%
  \BibitemOpen
  \bibfield  {author} {\bibinfo {author} {\bibfnamefont {V.~J.~M.}\ \bibnamefont {Lahuerta}}, \bibinfo {author} {\bibfnamefont {J.-N.}\ \bibnamefont {Kirsten-Siem{\ss}}}, \bibinfo {author} {\bibfnamefont {K.}~\bibnamefont {Hammerer}},\ and\ \bibinfo {author} {\bibfnamefont {N.}~\bibnamefont {Gaaloul}},\ }\href@noop {} {\bibinfo {title} {{Diffraction phase-free Bragg atom interferometry}}} (\bibinfo {year} {2025}),\ \Eprint {https://arxiv.org/abs/2505.23921} {arXiv:2505.23921} \BibitemShut {NoStop}%
\bibitem [{\citenamefont {Saywell}\ \emph {et~al.}(2023)\citenamefont {Saywell}, \citenamefont {Carey}, \citenamefont {Light}, \citenamefont {Szigeti}, \citenamefont {Milne}, \citenamefont {Gill}, \citenamefont {Goh}, \citenamefont {Perunicic}, \citenamefont {Wilson}, \citenamefont {Macrae} \emph {et~al.}}]{saywell2023enhancing}%
  \BibitemOpen
  \bibfield  {author} {\bibinfo {author} {\bibfnamefont {J.~C.}\ \bibnamefont {Saywell}}, \bibinfo {author} {\bibfnamefont {M.~S.}\ \bibnamefont {Carey}}, \bibinfo {author} {\bibfnamefont {P.~S.}\ \bibnamefont {Light}}, \bibinfo {author} {\bibfnamefont {S.~S.}\ \bibnamefont {Szigeti}}, \bibinfo {author} {\bibfnamefont {A.~R.}\ \bibnamefont {Milne}}, \bibinfo {author} {\bibfnamefont {K.~S.}\ \bibnamefont {Gill}}, \bibinfo {author} {\bibfnamefont {M.~L.}\ \bibnamefont {Goh}}, \bibinfo {author} {\bibfnamefont {V.~S.}\ \bibnamefont {Perunicic}}, \bibinfo {author} {\bibfnamefont {N.~M.}\ \bibnamefont {Wilson}}, \bibinfo {author} {\bibfnamefont {C.~D.}\ \bibnamefont {Macrae}}, \emph {et~al.},\ }\bibfield  {title} {\bibinfo {title} {Enhancing the sensitivity of atom-interferometric inertial sensors using robust control},\ }\href {https://doi.org/https://doi.org/10.1038/s41467-023-43374-0} {\bibfield  {journal} {\bibinfo  {journal} {Nat. Commun.}\ }\textbf {\bibinfo {volume} {14}},\ \bibinfo {pages} {7626} (\bibinfo
  {year} {2023})}\BibitemShut {NoStop}%
\bibitem [{\citenamefont {Rodzinka}\ \emph {et~al.}(2024)\citenamefont {Rodzinka}, \citenamefont {Dionis}, \citenamefont {Calmels}, \citenamefont {Beldjoudi}, \citenamefont {Beguin}, \citenamefont {Gu{\'e}ry-Odelin}, \citenamefont {Allard}, \citenamefont {Sugny},\ and\ \citenamefont {Gauguet}}]{rodzinka2024optimal}%
  \BibitemOpen
  \bibfield  {author} {\bibinfo {author} {\bibfnamefont {T.}~\bibnamefont {Rodzinka}}, \bibinfo {author} {\bibfnamefont {E.}~\bibnamefont {Dionis}}, \bibinfo {author} {\bibfnamefont {L.}~\bibnamefont {Calmels}}, \bibinfo {author} {\bibfnamefont {S.}~\bibnamefont {Beldjoudi}}, \bibinfo {author} {\bibfnamefont {A.}~\bibnamefont {Beguin}}, \bibinfo {author} {\bibfnamefont {D.}~\bibnamefont {Gu{\'e}ry-Odelin}}, \bibinfo {author} {\bibfnamefont {B.}~\bibnamefont {Allard}}, \bibinfo {author} {\bibfnamefont {D.}~\bibnamefont {Sugny}},\ and\ \bibinfo {author} {\bibfnamefont {A.}~\bibnamefont {Gauguet}},\ }\bibfield  {title} {\bibinfo {title} {{Optimal Floquet state engineering for large scale atom interferometers}},\ }\href {https://doi.org/https://doi.org/10.1038/s41467-024-54539-w} {\bibfield  {journal} {\bibinfo  {journal} {Nat. Commun.}\ }\textbf {\bibinfo {volume} {15}},\ \bibinfo {pages} {10281} (\bibinfo {year} {2024})}\BibitemShut {NoStop}%
\bibitem [{\citenamefont {Kotru}\ \emph {et~al.}(2015)\citenamefont {Kotru}, \citenamefont {Butts}, \citenamefont {Kinast},\ and\ \citenamefont {Stoner}}]{kotru2015large}%
  \BibitemOpen
  \bibfield  {author} {\bibinfo {author} {\bibfnamefont {K.}~\bibnamefont {Kotru}}, \bibinfo {author} {\bibfnamefont {D.~L.}\ \bibnamefont {Butts}}, \bibinfo {author} {\bibfnamefont {J.~M.}\ \bibnamefont {Kinast}},\ and\ \bibinfo {author} {\bibfnamefont {R.~E.}\ \bibnamefont {Stoner}},\ }\bibfield  {title} {\bibinfo {title} {{Large-area atom interferometry with frequency-swept Raman adiabatic passage}},\ }\href {https://doi.org/10.1103/PhysRevLett.115.103001} {\bibfield  {journal} {\bibinfo  {journal} {Phys. Rev. Lett.}\ }\textbf {\bibinfo {volume} {115}},\ \bibinfo {pages} {103001} (\bibinfo {year} {2015})}\BibitemShut {NoStop}%
\bibitem [{\citenamefont {Bongs}\ \emph {et~al.}(2006)\citenamefont {Bongs}, \citenamefont {Launay},\ and\ \citenamefont {Kasevich}}]{bongs2006high}%
  \BibitemOpen
  \bibfield  {author} {\bibinfo {author} {\bibfnamefont {K.}~\bibnamefont {Bongs}}, \bibinfo {author} {\bibfnamefont {R.}~\bibnamefont {Launay}},\ and\ \bibinfo {author} {\bibfnamefont {M.~A.}\ \bibnamefont {Kasevich}},\ }\bibfield  {title} {\bibinfo {title} {{High-order inertial phase shifts for time-domain atom interferometers}},\ }\href {https://doi.org/10.1007/s00340-006-2397-5} {\bibfield  {journal} {\bibinfo  {journal} {Appl. Phys. B}\ }\textbf {\bibinfo {volume} {84}},\ \bibinfo {pages} {599} (\bibinfo {year} {2006})}\BibitemShut {NoStop}%
\bibitem [{\citenamefont {Feynman}\ \emph {et~al.}(1957)\citenamefont {Feynman}, \citenamefont {Vernon~Jr},\ and\ \citenamefont {Hellwarth}}]{feynman1957geometrical}%
  \BibitemOpen
  \bibfield  {author} {\bibinfo {author} {\bibfnamefont {R.~P.}\ \bibnamefont {Feynman}}, \bibinfo {author} {\bibfnamefont {F.~L.}\ \bibnamefont {Vernon~Jr}},\ and\ \bibinfo {author} {\bibfnamefont {R.~W.}\ \bibnamefont {Hellwarth}},\ }\bibfield  {title} {\bibinfo {title} {{Geometrical representation of the Schr{\"o}dinger equation for solving maser problems}},\ }\href {https://doi.org/10.1063/1.1722572} {\bibfield  {journal} {\bibinfo  {journal} {J. Appl. Phys.}\ }\textbf {\bibinfo {volume} {28}},\ \bibinfo {pages} {49} (\bibinfo {year} {1957})}\BibitemShut {NoStop}%
\bibitem [{\citenamefont {Dick}(1989)}]{dick1989local}%
  \BibitemOpen
  \bibfield  {author} {\bibinfo {author} {\bibfnamefont {G.~J.}\ \bibnamefont {Dick}},\ }\bibfield  {title} {\bibinfo {title} {{Local oscillator induced instabilities in trapped ion frequency standards}},\ }in\ \href@noop {} {\emph {\bibinfo {booktitle} {Proceedings of the 19th Annual Precise Time and Time Interval Systems and Applications Meeting}}}\ (\bibinfo {year} {1989})\ pp.\ \bibinfo {pages} {133--147}\BibitemShut {NoStop}%
\bibitem [{\citenamefont {Cheinet}\ \emph {et~al.}(2008)\citenamefont {Cheinet}, \citenamefont {Canuel}, \citenamefont {Dos~Santos}, \citenamefont {Gauguet}, \citenamefont {Yver-Leduc},\ and\ \citenamefont {Landragin}}]{cheinet2008measurement}%
  \BibitemOpen
  \bibfield  {author} {\bibinfo {author} {\bibfnamefont {P.}~\bibnamefont {Cheinet}}, \bibinfo {author} {\bibfnamefont {B.}~\bibnamefont {Canuel}}, \bibinfo {author} {\bibfnamefont {F.~P.}\ \bibnamefont {Dos~Santos}}, \bibinfo {author} {\bibfnamefont {A.}~\bibnamefont {Gauguet}}, \bibinfo {author} {\bibfnamefont {F.}~\bibnamefont {Yver-Leduc}},\ and\ \bibinfo {author} {\bibfnamefont {A.}~\bibnamefont {Landragin}},\ }\bibfield  {title} {\bibinfo {title} {{Measurement of the sensitivity function in a time-domain atomic interferometer}},\ }\href {https://doi.org/10.1109/TIM.2007.915148} {\bibfield  {journal} {\bibinfo  {journal} {IEEE Trans. Instrum. Meas.}\ }\textbf {\bibinfo {volume} {57}},\ \bibinfo {pages} {1141} (\bibinfo {year} {2008})}\BibitemShut {NoStop}%
\bibitem [{\citenamefont {Templier}(2021)}]{templier2021three}%
  \BibitemOpen
  \bibfield  {author} {\bibinfo {author} {\bibfnamefont {S.}~\bibnamefont {Templier}},\ }\emph {\bibinfo {title} {{Three-axis hybridized quantum accelerometer for inertial navigation}}},\ \href@noop {} {Ph.D. thesis},\ \bibinfo  {school} {Universit{\'e} de Bordeaux} (\bibinfo {year} {2021})\BibitemShut {NoStop}%
\bibitem [{\citenamefont {Saywell}\ \emph {et~al.}(2021)\citenamefont {Saywell}, \citenamefont {Carey}, \citenamefont {Dedes}, \citenamefont {Kuprov},\ and\ \citenamefont {Freegarde}}]{saywell2021can}%
  \BibitemOpen
  \bibfield  {author} {\bibinfo {author} {\bibfnamefont {J.}~\bibnamefont {Saywell}}, \bibinfo {author} {\bibfnamefont {M.}~\bibnamefont {Carey}}, \bibinfo {author} {\bibfnamefont {N.}~\bibnamefont {Dedes}}, \bibinfo {author} {\bibfnamefont {I.}~\bibnamefont {Kuprov}},\ and\ \bibinfo {author} {\bibfnamefont {T.}~\bibnamefont {Freegarde}},\ }\bibfield  {title} {\bibinfo {title} {{Can optimised pulses improve the sensitivity of atom interferometers?}},\ }in\ \href {https://doi.org/https://doi.org/10.1117/12.2598991} {\emph {\bibinfo {booktitle} {Quantum Technology: Driving Commercialisation of an Enabling Science II}}},\ Vol.\ \bibinfo {volume} {11881}\ (\bibinfo {organization} {SPIE},\ \bibinfo {year} {2021})\ pp.\ \bibinfo {pages} {83--92}\BibitemShut {NoStop}%
\bibitem [{\citenamefont {Saywell}\ \emph {et~al.}(2022)\citenamefont {Saywell}, \citenamefont {Carey}, \citenamefont {Dedes}, \citenamefont {Kuprov},\ and\ \citenamefont {Freegarde}}]{saywell2022efficient}%
  \BibitemOpen
  \bibfield  {author} {\bibinfo {author} {\bibfnamefont {J.}~\bibnamefont {Saywell}}, \bibinfo {author} {\bibfnamefont {M.}~\bibnamefont {Carey}}, \bibinfo {author} {\bibfnamefont {N.}~\bibnamefont {Dedes}}, \bibinfo {author} {\bibfnamefont {I.}~\bibnamefont {Kuprov}},\ and\ \bibinfo {author} {\bibfnamefont {T.}~\bibnamefont {Freegarde}},\ }\bibfield  {title} {\bibinfo {title} {{Efficient state-symmetric beamsplitters and mirrors for atom interferometers using optimized pulses}},\ }\href {https://doi.org/https://doi.org/10.1088/1361-6455/ac8cc3} {\bibfield  {journal} {\bibinfo  {journal} {J. Phys. B: At. Mol. Opt. Phys.}\ }\textbf {\bibinfo {volume} {55}},\ \bibinfo {pages} {205501} (\bibinfo {year} {2022})}\BibitemShut {NoStop}%
\bibitem [{\citenamefont {Khaneja}\ \emph {et~al.}(2005)\citenamefont {Khaneja}, \citenamefont {Reiss}, \citenamefont {Kehlet}, \citenamefont {Schulte-Herbr{\"u}ggen},\ and\ \citenamefont {Glaser}}]{khaneja2005optimal}%
  \BibitemOpen
  \bibfield  {author} {\bibinfo {author} {\bibfnamefont {N.}~\bibnamefont {Khaneja}}, \bibinfo {author} {\bibfnamefont {T.}~\bibnamefont {Reiss}}, \bibinfo {author} {\bibfnamefont {C.}~\bibnamefont {Kehlet}}, \bibinfo {author} {\bibfnamefont {T.}~\bibnamefont {Schulte-Herbr{\"u}ggen}},\ and\ \bibinfo {author} {\bibfnamefont {S.~J.}\ \bibnamefont {Glaser}},\ }\bibfield  {title} {\bibinfo {title} {{Optimal control of coupled spin dynamics: design of NMR pulse sequences by gradient ascent algorithms}},\ }\href {https://doi.org/10.1016/j.jmr.2004.11.004} {\bibfield  {journal} {\bibinfo  {journal} {J. Magn. Reson.}\ }\textbf {\bibinfo {volume} {172}},\ \bibinfo {pages} {296} (\bibinfo {year} {2005})}\BibitemShut {NoStop}%
\bibitem [{\citenamefont {de~Fouquieres}\ \emph {et~al.}(2011)\citenamefont {de~Fouquieres}, \citenamefont {Schirmer}, \citenamefont {Glaser},\ and\ \citenamefont {Kuprov}}]{de2011second}%
  \BibitemOpen
  \bibfield  {author} {\bibinfo {author} {\bibfnamefont {P.}~\bibnamefont {de~Fouquieres}}, \bibinfo {author} {\bibfnamefont {S.~G.}\ \bibnamefont {Schirmer}}, \bibinfo {author} {\bibfnamefont {S.~J.}\ \bibnamefont {Glaser}},\ and\ \bibinfo {author} {\bibfnamefont {I.}~\bibnamefont {Kuprov}},\ }\bibfield  {title} {\bibinfo {title} {{Second order gradient ascent pulse engineering}},\ }\href {https://doi.org/10.1016/j.jmr.2011.07.023} {\bibfield  {journal} {\bibinfo  {journal} {J. Magn. Reson.}\ }\textbf {\bibinfo {volume} {212}},\ \bibinfo {pages} {412} (\bibinfo {year} {2011})}\BibitemShut {NoStop}%
\bibitem [{\citenamefont {Gershenzon}\ \emph {et~al.}(2008)\citenamefont {Gershenzon}, \citenamefont {Skinner}, \citenamefont {Brutscher}, \citenamefont {Khaneja}, \citenamefont {Nimbalkar}, \citenamefont {Luy},\ and\ \citenamefont {Glaser}}]{gershenzon2008linear}%
  \BibitemOpen
  \bibfield  {author} {\bibinfo {author} {\bibfnamefont {N.~I.}\ \bibnamefont {Gershenzon}}, \bibinfo {author} {\bibfnamefont {T.~E.}\ \bibnamefont {Skinner}}, \bibinfo {author} {\bibfnamefont {B.}~\bibnamefont {Brutscher}}, \bibinfo {author} {\bibfnamefont {N.}~\bibnamefont {Khaneja}}, \bibinfo {author} {\bibfnamefont {M.}~\bibnamefont {Nimbalkar}}, \bibinfo {author} {\bibfnamefont {B.}~\bibnamefont {Luy}},\ and\ \bibinfo {author} {\bibfnamefont {S.~J.}\ \bibnamefont {Glaser}},\ }\bibfield  {title} {\bibinfo {title} {{Linear phase slope in pulse design: Application to coherence transfer}},\ }\href {https://doi.org/https://doi.org/10.1016/j.jmr.2008.02.021} {\bibfield  {journal} {\bibinfo  {journal} {J. Magn. Reson.}\ }\textbf {\bibinfo {volume} {192}},\ \bibinfo {pages} {235} (\bibinfo {year} {2008})}\BibitemShut {NoStop}%
\bibitem [{\citenamefont {Koos}\ \emph {et~al.}(2015)\citenamefont {Koos}, \citenamefont {Feyrer},\ and\ \citenamefont {Luy}}]{koos2015broadband}%
  \BibitemOpen
  \bibfield  {author} {\bibinfo {author} {\bibfnamefont {M.~R.}\ \bibnamefont {Koos}}, \bibinfo {author} {\bibfnamefont {H.}~\bibnamefont {Feyrer}},\ and\ \bibinfo {author} {\bibfnamefont {B.}~\bibnamefont {Luy}},\ }\bibfield  {title} {\bibinfo {title} {{Broadband excitation pulses with variable RF amplitude-dependent flip angle (RADFA)}},\ }\href {https://doi.org/https://doi.org/10.1002/mrc.4297} {\bibfield  {journal} {\bibinfo  {journal} {Magn. Reson. Chem.}\ }\textbf {\bibinfo {volume} {53}},\ \bibinfo {pages} {886} (\bibinfo {year} {2015})}\BibitemShut {NoStop}%
\bibitem [{\citenamefont {Koos}\ \emph {et~al.}(2017)\citenamefont {Koos}, \citenamefont {Feyrer},\ and\ \citenamefont {Luy}}]{koos2017broadband2}%
  \BibitemOpen
  \bibfield  {author} {\bibinfo {author} {\bibfnamefont {M.~R.}\ \bibnamefont {Koos}}, \bibinfo {author} {\bibfnamefont {H.}~\bibnamefont {Feyrer}},\ and\ \bibinfo {author} {\bibfnamefont {B.}~\bibnamefont {Luy}},\ }\bibfield  {title} {\bibinfo {title} {{Broadband RF-amplitude-dependent flip angle pulses with linear phase slope}},\ }\href {https://doi.org/https://doi.org/10.1002/mrc.4593} {\bibfield  {journal} {\bibinfo  {journal} {Magn. Reson. Chem.}\ }\textbf {\bibinfo {volume} {55}},\ \bibinfo {pages} {797} (\bibinfo {year} {2017})}\BibitemShut {NoStop}%
\bibitem [{\citenamefont {Coote}\ \emph {et~al.}(2021)\citenamefont {Coote}, \citenamefont {Bermel},\ and\ \citenamefont {Arthanari}}]{coote2021optimization}%
  \BibitemOpen
  \bibfield  {author} {\bibinfo {author} {\bibfnamefont {P.}~\bibnamefont {Coote}}, \bibinfo {author} {\bibfnamefont {W.}~\bibnamefont {Bermel}},\ and\ \bibinfo {author} {\bibfnamefont {H.}~\bibnamefont {Arthanari}},\ }\bibfield  {title} {\bibinfo {title} {{Optimization of phase dispersion enables broadband excitation without homonuclear coupling artifacts}},\ }\href {https://doi.org/https://doi.org/10.1016/j.jmr.2021.106928} {\bibfield  {journal} {\bibinfo  {journal} {J. Magn. Reson.}\ }\textbf {\bibinfo {volume} {325}},\ \bibinfo {pages} {106928} (\bibinfo {year} {2021})}\BibitemShut {NoStop}%
\bibitem [{\citenamefont {Goodwin}\ \emph {et~al.}(2020)\citenamefont {Goodwin}, \citenamefont {Koos},\ and\ \citenamefont {Luy}}]{goodwin2020second}%
  \BibitemOpen
  \bibfield  {author} {\bibinfo {author} {\bibfnamefont {D.~L.}\ \bibnamefont {Goodwin}}, \bibinfo {author} {\bibfnamefont {M.~R.}\ \bibnamefont {Koos}},\ and\ \bibinfo {author} {\bibfnamefont {B.}~\bibnamefont {Luy}},\ }\bibfield  {title} {\bibinfo {title} {Second order phase dispersion by optimized rotation pulses},\ }\href {https://doi.org/https://doi.org/10.1103/PhysRevResearch.2.033157} {\bibfield  {journal} {\bibinfo  {journal} {Phys. Rev. Research}\ }\textbf {\bibinfo {volume} {2}},\ \bibinfo {pages} {033157} (\bibinfo {year} {2020})}\BibitemShut {NoStop}%
\bibitem [{Note1()}]{Note1}%
  \BibitemOpen
  \bibinfo {note} {Choosing to simulate an atomic momentum distribution with a $6\sigma $ width of $3\Omega _0$ in detuning means that 99.7\% of atoms will be within the optimization ensemble.}\BibitemShut {Stop}%
\bibitem [{\citenamefont {Bidel}\ \emph {et~al.}(2020)\citenamefont {Bidel}, \citenamefont {Zahzam}, \citenamefont {Bresson}, \citenamefont {Blanchard}, \citenamefont {Cadoret}, \citenamefont {Olesen},\ and\ \citenamefont {Forsberg}}]{bidel2020absoluteairborne}%
  \BibitemOpen
  \bibfield  {author} {\bibinfo {author} {\bibfnamefont {Y.}~\bibnamefont {Bidel}}, \bibinfo {author} {\bibfnamefont {N.}~\bibnamefont {Zahzam}}, \bibinfo {author} {\bibfnamefont {A.}~\bibnamefont {Bresson}}, \bibinfo {author} {\bibfnamefont {C.}~\bibnamefont {Blanchard}}, \bibinfo {author} {\bibfnamefont {M.}~\bibnamefont {Cadoret}}, \bibinfo {author} {\bibfnamefont {A.~V.}\ \bibnamefont {Olesen}},\ and\ \bibinfo {author} {\bibfnamefont {R.}~\bibnamefont {Forsberg}},\ }\bibfield  {title} {\bibinfo {title} {Absolute airborne gravimetry with a cold atom sensor},\ }\href {https://doi.org/10.1007/s00190-020-01350-2} {\bibfield  {journal} {\bibinfo  {journal} {J. Geod.}\ }\textbf {\bibinfo {volume} {94}},\ \bibinfo {pages} {20} (\bibinfo {year} {2020})}\BibitemShut {NoStop}%
\bibitem [{\citenamefont {Louchet-Chauvet}\ \emph {et~al.}(2011)\citenamefont {Louchet-Chauvet}, \citenamefont {Farah}, \citenamefont {Bodart}, \citenamefont {Clairon}, \citenamefont {Landragin}, \citenamefont {Merlet},\ and\ \citenamefont {Dos~Santos}}]{louchet2011influence}%
  \BibitemOpen
  \bibfield  {author} {\bibinfo {author} {\bibfnamefont {A.}~\bibnamefont {Louchet-Chauvet}}, \bibinfo {author} {\bibfnamefont {T.}~\bibnamefont {Farah}}, \bibinfo {author} {\bibfnamefont {Q.}~\bibnamefont {Bodart}}, \bibinfo {author} {\bibfnamefont {A.}~\bibnamefont {Clairon}}, \bibinfo {author} {\bibfnamefont {A.}~\bibnamefont {Landragin}}, \bibinfo {author} {\bibfnamefont {S.}~\bibnamefont {Merlet}},\ and\ \bibinfo {author} {\bibfnamefont {F.~P.}\ \bibnamefont {Dos~Santos}},\ }\bibfield  {title} {\bibinfo {title} {The influence of transverse motion within an atomic gravimeter},\ }\href {https://doi.org/https://doi.org/10.1088/1367-2630/13/6/065025} {\bibfield  {journal} {\bibinfo  {journal} {New J. Phys.}\ }\textbf {\bibinfo {volume} {13}},\ \bibinfo {pages} {065025} (\bibinfo {year} {2011})}\BibitemShut {NoStop}%
\bibitem [{\citenamefont {Peters}\ \emph {et~al.}(1999)\citenamefont {Peters}, \citenamefont {Chung},\ and\ \citenamefont {Chu}}]{peters1999measurement}%
  \BibitemOpen
  \bibfield  {author} {\bibinfo {author} {\bibfnamefont {A.}~\bibnamefont {Peters}}, \bibinfo {author} {\bibfnamefont {K.~Y.}\ \bibnamefont {Chung}},\ and\ \bibinfo {author} {\bibfnamefont {S.}~\bibnamefont {Chu}},\ }\bibfield  {title} {\bibinfo {title} {Measurement of gravitational acceleration by dropping atoms},\ }\href {https://doi.org/https://doi.org/10.1038/23655} {\bibfield  {journal} {\bibinfo  {journal} {Nature}\ }\textbf {\bibinfo {volume} {400}},\ \bibinfo {pages} {849} (\bibinfo {year} {1999})}\BibitemShut {NoStop}%
\bibitem [{\citenamefont {Templier}\ \emph {et~al.}(2021)\citenamefont {Templier}, \citenamefont {Hauden}, \citenamefont {Cheiney}, \citenamefont {Napolitano}, \citenamefont {Porte}, \citenamefont {Bouyer}, \citenamefont {Barrett},\ and\ \citenamefont {Battelier}}]{templier2021carrier}%
  \BibitemOpen
  \bibfield  {author} {\bibinfo {author} {\bibfnamefont {S.}~\bibnamefont {Templier}}, \bibinfo {author} {\bibfnamefont {J.}~\bibnamefont {Hauden}}, \bibinfo {author} {\bibfnamefont {P.}~\bibnamefont {Cheiney}}, \bibinfo {author} {\bibfnamefont {F.}~\bibnamefont {Napolitano}}, \bibinfo {author} {\bibfnamefont {H.}~\bibnamefont {Porte}}, \bibinfo {author} {\bibfnamefont {P.}~\bibnamefont {Bouyer}}, \bibinfo {author} {\bibfnamefont {B.}~\bibnamefont {Barrett}},\ and\ \bibinfo {author} {\bibfnamefont {B.}~\bibnamefont {Battelier}},\ }\bibfield  {title} {\bibinfo {title} {Carrier-suppressed multiple-single-sideband laser source for atom cooling and interferometry},\ }\href {https://doi.org/10.1103/PhysRevApplied.16.044018} {\bibfield  {journal} {\bibinfo  {journal} {Phys. Rev. Appl.}\ }\textbf {\bibinfo {volume} {16}},\ \bibinfo {pages} {044018} (\bibinfo {year} {2021})}\BibitemShut {NoStop}%
\bibitem [{\citenamefont {Skinner}\ \emph {et~al.}(2005)\citenamefont {Skinner}, \citenamefont {Reiss}, \citenamefont {Luy}, \citenamefont {Khaneja},\ and\ \citenamefont {Glaser}}]{skinner2005tailoring}%
  \BibitemOpen
  \bibfield  {author} {\bibinfo {author} {\bibfnamefont {T.~E.}\ \bibnamefont {Skinner}}, \bibinfo {author} {\bibfnamefont {T.~O.}\ \bibnamefont {Reiss}}, \bibinfo {author} {\bibfnamefont {B.}~\bibnamefont {Luy}}, \bibinfo {author} {\bibfnamefont {N.}~\bibnamefont {Khaneja}},\ and\ \bibinfo {author} {\bibfnamefont {S.~J.}\ \bibnamefont {Glaser}},\ }\bibfield  {title} {\bibinfo {title} {{Tailoring the optimal control cost function to a desired output: application to minimizing phase errors in short broadband excitation pulses}},\ }\href {https://doi.org/10.1016/j.jmr.2004.09.011} {\bibfield  {journal} {\bibinfo  {journal} {J. Magn. Reson.}\ }\textbf {\bibinfo {volume} {172}},\ \bibinfo {pages} {17} (\bibinfo {year} {2005})}\BibitemShut {NoStop}%
\bibitem [{\citenamefont {Hogben}\ \emph {et~al.}(2011)\citenamefont {Hogben}, \citenamefont {Krzystyniak}, \citenamefont {Charnock}, \citenamefont {Hore},\ and\ \citenamefont {Kuprov}}]{hogben2011spinach}%
  \BibitemOpen
  \bibfield  {author} {\bibinfo {author} {\bibfnamefont {H.~J.}\ \bibnamefont {Hogben}}, \bibinfo {author} {\bibfnamefont {M.}~\bibnamefont {Krzystyniak}}, \bibinfo {author} {\bibfnamefont {G.~T.}\ \bibnamefont {Charnock}}, \bibinfo {author} {\bibfnamefont {P.~J.}\ \bibnamefont {Hore}},\ and\ \bibinfo {author} {\bibfnamefont {I.}~\bibnamefont {Kuprov}},\ }\bibfield  {title} {\bibinfo {title} {{Spinach--a software library for simulation of spin dynamics in large spin systems}},\ }\href {https://doi.org/https://doi.org/10.1016/j.jmr.2010.11.008} {\bibfield  {journal} {\bibinfo  {journal} {J. Magn. Reson.}\ }\textbf {\bibinfo {volume} {208}},\ \bibinfo {pages} {179} (\bibinfo {year} {2011})}\BibitemShut {NoStop}%
\bibitem [{\citenamefont {Johansson}\ \emph {et~al.}(2012)\citenamefont {Johansson}, \citenamefont {Nation},\ and\ \citenamefont {Nori}}]{johansson2012qutip}%
  \BibitemOpen
  \bibfield  {author} {\bibinfo {author} {\bibfnamefont {J.~R.}\ \bibnamefont {Johansson}}, \bibinfo {author} {\bibfnamefont {P.~D.}\ \bibnamefont {Nation}},\ and\ \bibinfo {author} {\bibfnamefont {F.}~\bibnamefont {Nori}},\ }\bibfield  {title} {\bibinfo {title} {{QuTiP: An open-source Python framework for the dynamics of open quantum systems}},\ }\href {https://doi.org/https://doi.org/10.1016/j.cpc.2012.02.021} {\bibfield  {journal} {\bibinfo  {journal} {Comput. Phys. Commun.}\ }\textbf {\bibinfo {volume} {183}},\ \bibinfo {pages} {1760} (\bibinfo {year} {2012})}\BibitemShut {NoStop}%
\bibitem [{Note2()}]{Note2}%
  \BibitemOpen
  \bibinfo {note} {This definition is valid when the area under $h^{\triangle }(t)$ for the entire sequence produces an over-estimate of the true scale factor.}\BibitemShut {Stop}%
\end{thebibliography}%

\end{document}